\documentclass[article,showpacs,amsmath,amssymb,nofootinbib]{revtex4}

\usepackage[utf8]{inputenc}
\usepackage[english]{babel}
\usepackage{graphicx}
\usepackage{empheq}

\newcommand{\om} \omega   
\newcommand{\Om} \Omega
\newcommand{\eps} \epsilon

\newcommand{\be} {\begin{equation}}
\newcommand{\ee} {\end{equation}}
\newcommand{\ba} {\begin{eqnarray}}
\newcommand{\ea} {\end{eqnarray}}

\def\lrD{\mathrel{{\cal D}\kern-1.em\raise1.75ex\hbox{$\leftrightarrow$}}}
\def\lr #1{\mathrel{#1\kern-1.25em\raise1.75ex\hbox{$\leftrightarrow$}}}

\newcommand{\eqr}[1]{Eq.~(\ref{#1})}

\begin{document}

\title{From vacuum fluctuations across an event horizon \\to long distance correlations}

\author{Renaud Parentani}
\email{renaud.parentani@th.u-psud.fr}
\affiliation{Laboratoire de Physique Th\'eorique, CNRS UMR 8627, B\^at. 210, Universit\'e Paris-Sud 11, 91405 Orsay Cedex, France}

\date{\today}


\begin{abstract}

We study the stress energy  two-point function 
to show how 
short distance 
correlations across the horizon transform into 
correlations among 
asymptotic states, 
for the Unruh effect, and 
for black hole radiation. 
In the first case the transition is caused by the coupling to accelerated systems. 
In the second, the transition is more elusive and due to the 
change of the geometry from the near horizon region to the asymptotic one.
The gradual transition is appropriately described by 
using affine coordinates.  
We relate this to the covariant regularization used to evaluate the mean value of the stress energy. 
We 
apply these considerations to analogue black holes, i.e. dispersive theories. 
On one hand, the preferred rest frame gives further insight about the transition, 
and on the other hand, the dispersion tames the singular behavior 
found on the horizon in relativistic theories. 

\end{abstract}

\maketitle

\section{Introduction}

In Minkowski space 
the vacuum 
of quantum fields, the ground state of the Hamiltonian, 
is globally 
defined. 
For instance, for a massless free field $\phi$ in two 
dimensions,
it is the 
state annihilated by the destruction operators $a_k$ associated with the
positive norm 
modes $\phi_k = e^{-i\Om t +kz}/(4\pi \Om)^{1/2}$ where $\Om = |k| > 0$.
When probed locally, it determines 
correlation functions, such as that of the stress-energy tensor.
Considering 
$T_{UU} = (\partial_U \phi)^2$, 
where $U$ ($V$) is the retarded (advanced) null 
coordinate $U= t-z$, ($V= t+z$), the connected function is 
\ba
\langle T_{UU}({U}) \,  T_{UU}({U_0})\rangle &=& 
\Big(\partial_{{U}}\partial_{{U_0}}\,  \langle \phi({U}) \;  \phi({U_0})\rangle \Big)^2 
\nonumber\\
&=& 
\left(\frac{1}{4\pi}  \frac{1}{({U} - {U_0} - i \epsilon)^2}   \right)^2 .
\label{TUU}
\ea
The regulator $\epsilon \to 0^+$ specifies the nature
of the singularity in the 
limit ${U} \to {U_0}$. It arises from the fact
that only positive frequency $\Om = i\partial_{{U}}$ contribute to \eqr{TUU}.

Even though 
this 
function
monotonically behaves 
in $\Delta U^{-4}$,  
it determines 
subtle 
effects.
This
can 
be seen by considering 
accelerated systems where 
 both local and non-local phenomena
related to the Unruh 
effect~\cite{Unruh:1976db,Unruh:1983ms,Carlitz:1986ng,Massar:1995im,Brout:1995rd,Obadia:2002qe,Massar:2006ev}
can be obtained from \eqr{TUU}.
Moreover 
the correlations of $T_{UU}$ in the 
(Unruh)
 vacuum 
across the future horizon of a black hole behave as in \eqr{TUU}.
However since the black hole geometry is non uniform, as one probes the vacuum further away from the horizon, 
there is a gradual change 
which results, on one hand, in correlations typical of a thermal flux at 
the Hawking temperature~\cite{Hawking:1974sw,Davies:1976ei,Carlitz:1986ng} 
and, on the other, 
in 
correlations across the horizon between Hawking quanta and their partners~\cite{Massar:1996tx}.

The aim of the present paper is to analyze
the black hole 
case in the light of the Minkowski case. 
We show 
that the gradual change in the correlations
comes through the observables used 
to probe the state, and not from $\phi$ 
which is conformally invariant and thus insensitive to the 
geometry. We relate this gradual change 
to that obtained when considering 
renormalized expectation values which also depend on 
the conformal factor.
In the last part, 
we consider dispersive field theories~\cite{Unruh:1994je},
or analogue black holes in condensed matter~\cite{Unruh:1980cg,Barcelo}.
First, 
we show that
the preferred rest frame~\cite{Jacobson:1996zs} associated with dispersion defines 
new scalars (observables) 
which give an unambiguous meaning to the gradual change of 
the 
stress energy 
correlations.
Second we study how dispersion affects the correlation pattern.\footnote{
While completing this work, we became aware of \cite{Schutzhold:2010ig} where
similar conclusions are 
 reached.}
In addition to the robustness of the spectral properties of Hawking 
radiation~\cite{Unruh:1994je,Brout:1995wp,Corley96},
that  of the long distance correlations across the horizon 
was established
by constructing wave packets of $in$ modes~\cite{Brout:1995wp}.
We shall return to this observation and make contact with 
the analysis of density-density correlations in Bose-Einstein 
condensates~\cite{Balbinot:2007de,Iacopo08,Macher:2009nz}.

We have organized the paper as follows.
In Section II, we describe the behavior of the stress energy 2pt function in flat space,
and interpret  
its properties by introducing external 
(accelerated) systems which probe the field.
In Section III, we perform the same analysis in a black hole geometry.
Using affine coordinates, 
we present an invariant 
description of the transition from 
its near horizon behavior, which is similar to \eqr{TUU}, to its long distance behavior. 
Finally, in Section IV, 
we consider the stress energy correlations in 
dispersive theories, 
and 
investigate what are the modifications 
induced by dispersion. We conclude with 
remarks about the similarities between these dispersive effects and those
obtained with relativitic fields propagating in fluctuating metrics~\cite{Barrabes:2000fr,
Parentani:2007mb}.



\section{Unruh effect and correlations across a Rindler horizon}

We introduce the null coordinates $a u= - \ln(-aU)$, $a v = \ln(aV) $,  defined
respectively for $U< 0$ and $V > 0$. 
These are related by $\tau = (u+v)/2$
to $\tau$, the proper time of an accelerated system which follows 
$z^2 - t^2 = 1/a^2$ in $R$, the right quadrant $z > |t|$. ($a$ is the constant acceleration.)
%
When ${U}$ and ${U_0}$ are 
negative, i.e. with both points on the right of $U=0$, the future horizon 
 of the accelerated system, 
\eqr{TUU} 
can be re-written using $u$. 
%
Introducing $T^R_{uu} = (\partial_u \phi)^2 = T_{UU} \, e^{- 2 a u}$, \eqr{TUU} becomes 
\ba
\langle T^R_{uu}({u}) \,  T^R_{uu}({u_0})\rangle &=& \left(\frac{1}{16 \pi} 
\frac{a^2}{( \sinh \frac{a}{2}({u} - {u_0} - i\epsilon))^2}\right)^2 .
\label{Tuu}
\ea
Since $1/(\sinh y)^2= \Sigma_n\,  (y + i n \pi)^{-2}$,
where the sum over $n$ 
goes from $-\infty$ to $\infty$, 
$\langle T^R_{uu}  T^R_{uu}\rangle $ 
 is {\it the} periodic function in Im$u$ 
of period $2\pi/a$ whose zeroth term, $n=0$, 
behaves as in the vacuum, see \eqr{TUU}. 
Hence \eqr{Tuu} is 
the 2-point 
function 
in a thermal bath 
at a temperature equal to $a/2\pi$.
Noticing that 
along the accelerated trajectory, one has $\tau = u$ (because $v=u$)
\eqr{Tuu}
determines 
the 
correlation function of the 
energy density 
probed 
by the accelerated system. 
In this way we recover 
the Unruh effect: accelerated systems perceive the vacuum as a thermal ensemble at that 
temperature~\cite{Unruh:1976db}.


It is less usual but equally
interesting to consider \eqr{TUU} with one point on either side of the horizon. 
Using $u$ and $T_{uu}$ on the right,
and their symmetrical counterpart on the left, 
$a \bar u = \ln(aU)$
and $T^L_{u u} = (\partial_{\bar u} \phi)^2 = T_{UU} \, e^{2 a \bar u}$ for $U > 0$, 
\eqr{TUU} becomes 
\ba
\langle T^L_{u u}(\bar {u}) \,  T^R_{uu}({u_0})\rangle &=& \left(\frac{1}{16 \pi}
\frac{ a^2}{(\cosh \frac{a}{2}(\bar {u} + {u_0}))^2}\right)^2  .
\label{ldTuu}
\ea 
Quite 
surprizingly  
$\langle T^L_{u u} T^R_{uu}\rangle$ possesses a 
'bump'
which is centered at $\bar {u} + {u_0} = 0$.
In terms of the affine coordinate $U$, 
it is located at ${U} = - {U_0}$, 
on the opposite value from the horizon. 
This 
is unnexpected 
because, 
at {\it fixed} $\Delta V$, $\langle \phi(U, V) \phi(U_0,V_0) \rangle 
 = \frac{-1}{ 4\pi} \ln \Delta U
 + Cst. $ 
 has no maximum 
when expressed in terms of $\bar u$ and $u_0$. 
In fact, the 
maximum in \eqr{ldTuu} 
results from the combined effect of the Jacobians $dU/du$ 
associated with the tensorial character of $T_{UU}$,  
and the monotonic power law decay of \eqr{TUU}.\footnote{My attention was drawn to this point
in a discussion with W. Unruh and T. Jacobson that took place during 
the workshop “Towards the observation of
Hawking radiation in condensed matter systems” 
held at IFIC in Valencia in Feb. 2009.} 
These 
properties 
apply to massless fields in any space-time dimension $d$. Because of the dimensionality of the $\phi$ field
($= (d-2)/2$) 
the location of the maximum is shifted at 
$\bar {u} + {u_0} = -\frac{2}{a} {\rm arctanh} \frac{d-2}{d+2}.$


So far however
\eqr{ldTuu} is simply a 
re-expression of \eqr{TUU}: 
no physics is gained when writing a tensor in a new coordinate system.
So, isn't the maximum of \eqr{ldTuu} just a coordinate artefact ?
%
No, \eqr{ldTuu}, as \eqr{Tuu}, governs physical effects. 
For instance, when looking at the 
state of two opposite 
accelerated systems~\cite{Massar:2006ev}, 
the correlation matrix of this bi-partite system 
exhibits a non-trivial entanglement  near  
the maximum of \eqr{ldTuu}. 
Similarly the scattering by two opposite 
accelerated mirrors destructively interferes near that maximum~\cite{Obadia:2002qe}.
As another example~\cite{Massar:1995im}, 
expressions closely related to \eqr{ldTuu}
obtain when looking at the 
value of $T^L_{uu}$ which is correlated 
to the fact that an accelerated detector in $R$ has 
made a transition. 
In all cases the physics is the same: 
measurements performed in  $R$ 
are statistically correlated to those 
performed on the other side of the horizon.

To convey the idea that 
the re-writing of \eqr{TUU} as Eqs. (\ref{Tuu}, 
\ref{ldTuu}) is not merely a 
coordinate change 
but has to do with the physics of accelerated systems,
we make two observations. First 
it is conceptually useful 
to 
introduce the ($U$ part of the) {\it scalar} 
energy density that such system in $R$ will measure: 
$\rho^R = T_{\mu \nu} u_{R}^\mu  u_{R}^\mu $,
where $u_{R}^\mu= dx_{R}^\mu/d\tau$ is the unit vector field tangeant to 
its trajectory. Then the l.h.s of \eqr{Tuu} is 
$\langle \rho^R 
(\tau) \, \rho^R 
(\tau_0)
\rangle$.
Similarly the l.h.s of \eqr{ldTuu} is 
$\langle \rho^L(\bar\tau) \, \rho^R 
(\tau_0) \rangle$.
Hence both 
\eqr{Tuu} and \eqr{ldTuu} 
have a clear 
interpretation
as energy-energy correlation functions of accelerated systems.
Secondly, we 
notice that 
\eqr{Tuu} only depends on ${u} - {u_0}$.
This 
results from the invariance of the vacuum under 
Lorentz transformations 
$u \to u + b, U \to U e^{-ab}$.  
In fact the 
Minkowski 
distance between two points situated along the accelerated trajectory $v=u$ 
is
\be
s^2 = - \Delta t ^2 + \Delta z ^2 
= - \frac{4 \sinh^2(a (\tau- \tau_0)/2)}{a^2}.
\label{s1}
\ee
We therefore see that the stationarity 
of \eqr{Tuu} in $u$ for {\it arbitrary values}  of $v$
are related to the properties of $s^2(\Delta\tau)$ evaluated along $v=u$.
In a similar fashion, \eqr{ldTuu} is related to 
the 
distance $s_{\rm op}^2$ bewteen two points situated along the 
opposite accelerated trajectories defined by $z^2 - t^2 = 1/a^2$. Using the proper time $\tau_0$ ($\bar \tau$)
to localize the point on the right (left) trajectory, one gets
\be
s_{\rm op}^2 = 
\frac{4 \cosh^2(a (\bar \tau + \tau_0)/2)}{a^2}.
\label{s2}
\ee
The maximum of correlation in \eqr{ldTuu} thus coincides with 
the minimum (space-like) distance bewteen the two 
trajectories. 
 Because of Lorentz invariance no
preferred spatial distance $\Delta z ^2$ is selected though.

Yet, in preparation for the black hole problem,  
one would like to know what is the origin of 
the {\it mathematical} properties of \eqr{Tuu} and \eqr{ldTuu}.
In fact,  these 
must be deeply rooted
since 
\eqr{ldTuu} is the analytical continuation of \eqr{Tuu} obtained by 
subtracting 
half of the imaginary 
period 
that fixes the Unruh 
temperature. 
In addition, we wish to identify what could explain 
the maximum of correlation in \eqr{ldTuu}, 
which is independent of 
$\Delta V$, and thus 
{not necessarily} refers to \eqr{s2} and 
 accelerated systems in $L$. 
As shown in the Appendices, 
these 
properties can be understood 
when using 
the 
Unruh modes~\cite{Unruh:1976db}. 
In brief, the maximum of \eqr{ldTuu} finds its origin in the 
entanglement in Fock space between states with opposite Killing frequency $\om = i\partial_u$. 
This entanglement translates into space-time correlations because, in each pair,
the partner mode lives on the other side of the Rindler horizon, {\it and} is weighted in \eqr{Um}
by a factor $z_\om$ which is real for all $\om$.
In addition the  maximum value of correlations is directly related to the mean occupation
number of Rindler quanta. 

\section{Hawking effect and
correlations across the 
 horizon}


We have a double aim. First, we wish to interpret the equivalent of \eqr{TUU}
and 
Eqs. (\ref{Tuu},\ref{ldTuu}) in a black hole geometry.
Second, we aim to 
analyze the differences 
 between the black hole case and Minkowski.
In particular we are looking for a smooth and {\it coordinate invariant} interpolation from the near horizon region,
where \eqr{TUU} should make sense, to the asymptotic regions where Eqs. (\ref{Tuu},\ref{ldTuu}) 
should do.

To reach these ends, 
we 
need to identify what corresponds to  
the Minkowski vacuum, 
the affine coordinate $U$, and 
the 
$u$  coordinate 
associated 
with boosts 
$u \to u+b, U \to U e^{ab}$. 
For simplicity we shall work with a two dimensional conformally invariant field
 $\phi$. 
Then the identification is rather easy and known~\cite{Unruh:1976db,Brout:1995rd}.
The novelty mainly consists in the attention paid 
to the stress-energy correlation function.


\subsection{Prelude: conformal invariance}

We 
work with a 
conformally invariant field 
because 
the simplicity of the expressions 
will oblige us to identify under which conditions 
identities 
valid in all space-times
acquire a 
physical meaning related to the Unruh and/or to the Hawking effect. 

In 
two dimensions 
the line element can always be written in double null coordinates as 
\be
ds^2 = - C(U,V)\,  dU dV \, .
\ee
In these coordinates, 
$\phi$ obeys 
$\partial_U \partial_V \phi = 0$. 
Therefore the conformal factor 
$C$
drops out and the right and left moving sectors 
remain 
decoupled $\phi = \phi(U) + \phi(V)$. 
As a result, for {any} $U$ coordinate, one can consider the $U$-vacuum state 
defined by the positive frequency modes $\phi_\Om = e^{-i \Om U}/(4 \pi \Om)^{1/2}$.
In that state, the {\it connected}
correlation function of
$T_{UU} = (\partial_U \phi)^2$
\be
\langle T_{UU}({U}) \,  T_{UU}({U_0})\rangle_{c} 
\equiv
 \langle T_{UU}({U}) \,  T_{UU}({U_0})\rangle-  \langle T_{UU}({U}) \rangle \,   \langle T_{UU}({U_0}) \rangle,
\label{TUUC}
\ee
obeys \eqr{TUU}, 
independently 
of the regularization scheme used to compute $\langle T_{UU}({U}) \rangle$.\footnote{
The 
connected 2pt function of $T_{\mu \nu}$
governs gravitational back-reaction effects 
beyond those included in the semi-classical Einstein equations.~\cite{Parentani:2007mb}
This provides its dynamical relevance.} 
Introducing the coordinates $au =\ln( -aU)$
and $a \bar u = \ln (aU)$ for respectively negative and positive values of $U$, \eqr{Tuu} and \eqr{ldTuu} 
also follow, as 
{\it mathematical identities} valid in all space-times and for any coordinate $U$.
%
Some physical input is therefore needed 
to transform these identities into meaningful relations among observables.

To relate \eqr{Tuu} 
to the Unruh effect is straightforward. It suffices to consider
a 
particle detector following the orbit $u=v$. It will perceive the $U$-vacuum
as a thermal state at a temperature $a/2\pi$. (In general $a$ is no longer the proper acceleration, 
consider e.g.  
de Sitter space where the thermal bath is also perceived by inertial detectors.) Similarly, for \eqr{ldTuu}, 
when introducing another detector which follows 
the mirror trajectory $\bar u = \bar v$, 
the combined state of these two detectors will be entangled as that of accelerated detectors
in Minkowski~\cite{Massar:2006ev}, again as a direct consequence
of the conformal invariance of $\phi$.
This generalization of the Unruh effect 
applies to any 
field and in any dimension, but only approximatively provided the acceleration is much higher 
that the space-time curvature. 
(This 
limit is also used 
in the thermodynamic analysis of space-time~\cite{Jacobson:1995ab}. 
However the fact that \eqr{ldTuu} always applies 
indicates that the purity of the quantum state, and not only the mean fluxes, is 
a key ingredient.)


To relate 
\eqr{Tuu} and \eqr{ldTuu} to the Hawking effect is more elaborate. 
As we shall see, the metric 
plays a crucial role in 
"selecting" the $U$-vacuum and the coordinate $u$. 
This follows from 
the fact that, unlike the connected 
2pt function of \eqr{TUUC}, 
other 
observables are {\it not} conformally invariant, 
as for
instance the renormalized 
expectation value of $T_{UU}$~\cite{Davies:1976ei}
\be
\langle T_{UU}\rangle^{\rm ren} = - \frac{1}{12 \pi}  C^{1/2} \, \partial_U^2 \frac{1}{C^{1/2}} . 
\label{Tuuren}\ee
In Sec. III.D, we shall see that 
the breakdown of the conformal invariance comes from the 
renormalization scheme and not from the "bare" operator.

\subsection{The correspondence}

We work with one dimensional 
 stationary  black hole metrics that we describe using
Eddington-Finkelstein (EF) coordinates $v,r$ 
in which the line element reads
\ba
ds^2 &=& -(1- w^2(r))\, dv^2 + 2 dv dr .
\label{EF} 
\ea
We shall not consider specific functions for $w^2$, 
as e.g. $2M/r$ which describes a 
Schwarzschild black hole. 
Hence the coordinate $r$ should not be 
conceived as associated
with spherical symmetry. In fact, from a 2D point of view, 
$r$ in \eqr{EF}
is an affine coordinate 
along $v=cst.$ We also notice that $r$ is affine along $\tau = cst.$, 
where $\tau$ is the proper time appearing in Painlev\'e-Gullstrand
(PG) coordinates~\cite{Barcelo} where $ds^2 = - d\tau^2 + (dr - w d\tau )^2$.
This guarantees that the forthcoming analysis can be made in PG coordinates.

When $w$ is 
a constant, the geometry is 
Minkowski.  
Therefore the differences with 
Sec. II will {\it all} stem from
the gradient of 
$w(r)$.
We have adopted these coordinates 
because they are well behaved
in the late time portion of the space-time
which is relevant for the Hawking effect and the correlations across the horizon, see e.g.  \cite{Brout:1995rd}.
In particular they stay regular 
when $w^2$ crosses $1$, say at $r= r_h $.
This locus is the future horizon (for the static observers at fixed $r > r_h$) 
and it is a null line $U= cst$. 
The absence of the corresponding $V$ (past) horizon
induces a 
disymmetry:
left moving modes 
are regular across the horizon $U=cst.$ and 
play 
no role in the following. 

Assuming that $w(r)\to w_\infty = cst.$ for $r \to \infty$, 
the asymptotic observers at fixed $r$ are inertial, at rest with the black hole, and 
their {proper} time is 
$\tau_{\rm as} = v 
(1 - w_\infty^2)^{1/2}$. 
By rescaling 
$v$ and $r$, one can 
work in a gauge in which $w_\infty= 0$. 
From now on we adopt it. 
The positive norm modes that these observers will use 
are 
$ \phi_\om^{\rm as} = {e^{- i\om u } / 
(4\pi \om)^{1/2}},
$ 
with $\om > 0$.
When $r \to \infty$, 
 $u,v$ are related to 
$\tau_{as}$ by $\tau_{as} = (u+v)/2$. 
At finite $r$,
one has $u 
= v - 2 \int dr/(1-w^2)$.
%
One sees that
$u$ might diverge when $w^2$ crosses $1$. 

To settle this, 
one needs to know how $w$ behaves across $r_h$. We assume a regular behavior 
$w^2(r) \sim  1 - 2  \kappa (r - r_h)$ 
as this is 
the case for regular collapses. ($\kappa > 0$ coincides with the 
surface gravity since 
$w_\infty = 0$~\cite{Jacobson:2007jx}. The extreme case $\kappa = 0$ requires a special 
treatment~\cite{Balbinot:2007kr}.) 
At fixed $v$, 
one finds $u \sim - \kappa \ln(r- r_h)$.
Therefore, 
when 
$r \to r_h^+$ at fixed $v$,
$\phi^{\rm as}_\om$ 
behaves as 
\be
\phi^{\rm as}_\om(u) \sim
{(r_v(u)-r_h)^{i\om/ \kappa} \over (4\pi \om)^{1/2}}
, 
\label{bhm}
\ee
and can be taken to vanish for $r< r_h$. The function $r_v(u)$ obeys 
\be
{dr_v} = -\frac{1}{2} \, C(u,v) du ,
\label{rv}
\ee
where $C(u,v)= (1- w^2)$ is the conformal factor in the $u,v$ coordinates. 
It should be emphasized that $r_v(u)$ is an affine parameter for all $w(r)$, 
and this property is governed by 
$C$ which appears here for the second time.
(It was implicitly used before 
when imposing that the coordinates $u, v$ are related to the asymptotic
proper time by $2\tau_{as} = u+v$.) 

When replacing the acceleration $a$ by $ \kappa$ in \eqr{bhm}, the correspondance with 
 \eqr{Um} is manifest, and  
physically 
meaningful 
 because 
$r_v$ 
is affine,
as  $U$ is 
in Minkowski space. 
However, this correspondence is confined near the horizon 
since 
$\phi^{\rm as}_\om$
behaves as $e^{i 2\om r}$ for $r\to \infty$ at fixed $v$.
Therefore, the gradual change of the mode $\phi_\om$ 
from
$\sim (r-r_h)^{i\om/\kappa}$ to 
$\sim e^{i 2\om r}$ is what distinguishes
the back hole case from Minkowski. These 
facts have been recognized in~\cite{Unruh:1976db,Damour:1976jd} and shall be further exploited 
below.

The second part of the correspondence concerns the state of $\phi$.
While in Minkowski space the notion of vacuum is unambiguous, 
in a black hole geometry 
this 
is lost. Nevertheless 
the late time behavior~\cite{Hawking:1974sw}
is universal {\it and} stationary when described in terms of $u$ (or $\tau_{\rm as}$). 
This can be 
verified by considering different collapses and different initial states:
in each case there are transients, 
but after a few e-folding $u$-times
$1/\kappa$, these fade out as 
$\sim e^{-\kappa u}$
and the stationary values set in
(unless the collapse and/or the state is singular).
Therefore, as far as the description in terms of $u$ 
is concerned,
a {\it single} stationary state 
is 
selected. 
As 
noticed in~\cite{Unruh:1976db}, this state is most simply characterized 
by Unruh 
modes,  $\phi_\om$, 
exactly as the Minkowski vacuum can also be, see the discussion after \eqr{Um}.
The particular combination of modes 
weighted by $z_\om = e^{- \pi \om/\kappa}$ means that $\phi_\om$ only contains positive frequencies 
$\Omega_K= i\partial_{U_K}$
where $U_K$ is {\it the} regular 
coordinate 
across $r_h$ which is related to $u$ by 
$d U_K/du = e^{-\kappa u}$.
Indeed, stationarity requires an exponential relation bewteen $u$ and $U_K$,
whereas regularity across $r_h$ fixes the decay rate $d \ln U_K /du$ 
to be 
$\kappa$.

The correspondence is now completed: 
the regularity and the stationarity of the state 
allow to characterize it 
by the modes
 of 
\eqr{Um}, with 
 $U$ replaced by 
$U_K 
= -e^{-\kappa u}/\kappa$. 
This exponential 
is no longer related to a Lorentz transformation,
but is still related to an isometry whose
Killing field norm $\propto (1 - w^2)^{1/2}$ 
vanishes on the horizon.

\subsection{Correlation functions and asymptotic quanta}

In the Unruh vacuum,  
the connected 2pt function  of  $T_{UU}= (\partial_{U_K} \phi)^2$, see \eqr{TUUC},
is {exactly} 
given by \eqr{TUU}. 
Then, 
since 
$\kappa u = -\ln (-\kappa U_K)$,
\eqr{Tuu} with $a \to \kappa$ 
also obtains. 
However, as discussed in III. A, there is nothing special about these identities. 
What 
makes 
\eqr{Tuu} 
meaningfull here is 
that both $U_K$ and $u$ are affine, respectively across the horizon and asymptotically.
Hence \eqr{Tuu} tells us that
when evaluated in the {regular} 
state,
the 2pt function used by {inertial} asymptotic observers 
 is that of 
a thermal flux at the Hawking temperature $\kappa/2\pi$~\cite{Carlitz:1986ng}.

Let us now consider the equivalent of \eqr{ldTuu}. Introducing on the other side of the horizon
$\kappa \bar u = \ln \kappa U_K$, 
 \eqr{ldTuu} with $a \to \kappa$ automatically obtains. But what does it mean~?
Unlike for the Unruh effect, we cannot refer to accelerated systems in the $L$ quadrant.
Nevertheless the procedure of \cite{Massar:2006ev} applies to black holes,
and one can study the $T_{\mu \nu}$ correlated to the detection of an asymptotic quantum.
As in Minkowski, 
there is a reduction of the state:
 expectation values 
should be computed with the reduced density matrix, see App.~C. 
Then, because of the entanglement of \eqr{sqs},  
there is a correlation between $T_{\mu \nu}$ evaluated inside the horizon and a detection 
on $\cal I^+$~\cite{Massar:1996tx}. 
Equivalently, 
one can directly look for correlations in energy across the horizon and obtain 
\eqr{ldTuu}.

There is however an important difference between \eqr{Tuu} and \eqr{ldTuu}. It
originates from the different status of the coordinates $u$ and $\bar u$.
\eqr{Tuu} has a clear interpretation because
$du $ at fixed $v$ is affine for $r\to \infty$.
If we can 
 consider 
the equivalent of \eqr{Tuu} with both points inside, and obtain the same answer when using $\bar u$,
we should ask under which conditions would $\langle T^L_{uu} T^L_{uu}\rangle$ posses 
an intrinsic meaning (without any reference to the external region). 
For this, it "suffices" that the inside region be also infinite 
and that $w \to const.$ for $r \to - \infty$. Whereas it is unlikely that this be relevant for
black holes, 
we assume it is the case 
and consider the consequences. 
(In a next Section we discusss analogue black holes 
where this possibility can easily be realized.)
%
When $w \to const.$ for $r \to - \infty$, 
the notion of asymptotic quanta equally applies to the negative frequency partners~\cite{Macher:2009tw}.
Then, $\langle T^L_{uu} T^L_{uu}\rangle$ 
has 
the same meaning as $\langle T^R_{uu} T^R_{uu}\rangle$,
and
the long distance correlation $\langle T^L_{uu} T^R_{uu}\rangle$ of \eqr{ldTuu}
can be probed.
%



\subsection{Locality, covariance and renormalization}

To interprete \eqr{TUUC} 
we have 
so far used the asymptotic properties of the metric. 
However we would like an {intrinsic} description of the 
{\it gradual} 
transition from the horizon 
to the asymptotic regions. 
To this end we consider the renormalization procedure.
As we shall see its 
covariance 
supplies the intrinsic description we are looking for, and this by breaking the conformal invariance. 
(It should be clear 
that different rules 
on how to interpolate 
will give different behaviors since the 2pt function of $T_{UU}$ 
is not a scalar).

We noticed 
that both $u$ and $U_K$
are 
affine 
respectively for $r \to \infty$ and $r \sim r_h$.  
We also noticed that $r_v$ of \eqr{rv} is affine 
all the way through. 
In what follows we exploit this 
%
to relate the 2pt 
function of $T_{UU}$ to its {renormalized} value of \eqr{Tuuren}. 
To this end, 
we study the coincidence point limit of \eqr{TUUC} in the Unruh vacuum
using 
\be
T_{rr} = (\partial_r \phi)^2 =  \left(\frac{dU}{dr}\right)^2\,  T_{UU} ,
\label{Trr}
\ee which 
is "coordinate invariant" since $r_v$ is,  up to a scale, 
globally defined. 
To present the concepts we first work at spatial infinity. There, in the limit $u_0 \to u$, one has 
\ba
\langle T_{uu}(u) T_{uu}(u_0)\rangle_{K} &=&  
\left( \frac{dU_K(u)}{du}  \frac{dU_K(u_0)}{du_0} \right)^2
\left( 
\frac{1}{4\pi}\frac{1}{({U_K(u)} - {U_K(u_0)} - i \epsilon)^2}   \right)^2
\nonumber \\
&=&  
\left(\frac{1}{4\pi}  \frac{1}{({u} - {u_0} - i \epsilon)^2}   \right)^2 
+ 
\frac{ 2\, \langle T_{uu}(u)\rangle^{\rm ren}_{K} }{4\pi({u} - {u_0} - i \epsilon)^2} 
+ O\left(\frac{1}{u - u_0}\right), 
\label{TTrenas}
\ea
where $\langle T_{uu}(u)\rangle^{\rm ren}_{K}$ is the asymptotic 
expectation value of $T_{uu}$ 
in the Unruh vacuum. Indeed, it is 
defined~\cite{Davies:1976ei,Brout:1995rd} as 
\ba
\langle T_{uu}(u)\rangle^{\rm ren}_{K} &=& \lim_{u_0 \to u} \partial_u \partial_{u_0}
 \Big(\langle \phi(u) \phi(u_0)\rangle_{K} - \langle \phi(u) \phi(u_0)\rangle_M \Big)
\nonumber \\
&=& \frac{1}{12 \pi} 
\left( \frac{dU_K}{du} \right)^{1/2}  \partial_{u}^2 
\left( \frac{dU_K}{du} \right)^{-1/2}
= \frac{\kappa^2}{48\pi},
\label{Tuuas}
\ea
where $\langle \phi(u) \phi(u_0)\rangle_M = \frac{-1}{4\pi}\ln ({u} - {u_0} - i \epsilon)$ is 
evaluated in the asymptotic (Boulware) Minkowski-like vacuum.
The procedure of \eqr{TTrenas} is clear: by subtracting the asymptotic vacuum divergence, 
one extracts the excess in the "noise" and thus identifies
the mean value of \eqr{Tuuas}.
Moreover, this procedure can be {\it univocally} covariantized and applied 
at every space time point. 
This is 
achieved 
by making use of $r_v(u)$:
\ba
\langle T_{rr}(u)\,  T_{rr}(u_0)\rangle_{K} &=&  
\left(\frac{1}{4\pi} 
 \frac{1}{(r_{v}{(u)} - r_v{(u_0)} + i \epsilon)^2}   \right)^2
 \nonumber \\
&& + 2 \, {\langle T_{rr}(u,v)\rangle^{\rm ren}_{K} } 
\, \left(\frac{1}{4\pi} 
 \frac{1}{(r_{v}{(u)} - r_v{(u_0)} + i \epsilon)^2}   \right) + ...
\label{loccol}
\ea
where $\langle T_{rr}(u,v)\rangle^{\rm ren}_{K}$ is the 
renormalized value of $T_{rr}$ 
which is 
defined as 
\ba
\langle T_{rr}(u,v)\rangle^{\rm ren}_{K}
&=& \lim_{u_0 \to u} \partial_r \partial_{r_0}
 \Big(\langle \phi(u) \phi(u_0)\rangle_{K} - \langle \phi(u) \phi(u_0)\rangle_{\rm local}
\Big)\nonumber \\ 
&=& 
\frac{1}{12 \pi} 
\left( \frac{dU_K}{dr} \right)^{1/2}  \partial_{r}^2 
\left( \frac{dU_K}{dr} \right)^{-1/2},
\label{Tuuloc}
\ea
where 
\be
\langle \phi(u) \phi(u_0)\rangle_{\rm local} = - \frac{1}{4\pi}
\ln (r_v(u) - r_v(u_0) + i \epsilon). 
\label{localG}
\ee
In virtue of the covariance, when using 
the affine 
coordinate $r_v$, the subtraction term 
possesses this universal form.
This explains why \eqr{Tuuloc} 
generalizes \eqr{Tuuas} at every point. 

Several remarks should be made.
Firstly, using \eqr{rv}, one verifies that \eqr{Tuuloc} gives back \eqr{Tuuren} 
in all space-times and all vacua. Hence \eqr{Tuuloc} can be seen as an alternative  
expression for it. 
Secondly, since 
the vacuum 
is 
defined through the coordinate $U_K$, and the subtraction only refers to the metric through 
\eqr{rv}, 
$\langle T_{rr}\rangle^{\rm ren}_{K}$ is 
governed by $dU_K/dr$ and nothing else.  
%
Thirdly, 
even though $\langle T_{UU} T_{UU}\rangle_K$ is independent of 
$v$, 
the covariance of the divergent terms in \eqr{loccol} 
unambiguously 
defines the $v$-dependence 
of 
$\langle T^{\rm ren}_{rr}\rangle_K$ in \eqr{Tuuloc}.

\subsection{Covariant description of stress tensor correlations}

Following the same logic, we use 
$T_{rr}$ of \eqr{Trr}
to characterize the 
stress-energy correlations 
in the Unruh vacuum at every point. Using \eqr{TUU}, \eqr{TUUC} gives
\ba
\langle T_{rr}(v,r)\,  T_{rr}(v_0 , r_0)\rangle_{K} &=&
\Big(  \partial_{r} \partial_{r_0}
\langle \phi(v,r)\, \phi(v_0,r_0)\rangle_{K} \Big)^2
\nonumber\\
&=& 
\left( \frac{dU_K}{dr}  \frac{dU_K}{dr_0} \right)^2 \left( 
\frac{1}{4\pi}\frac{1}{({U_K(v,r)} - {U_K(v_0,r_0)} - i \epsilon)^2 }  \right)^2 .
\label{TTrr}
\ea
This bi-tensor field 
depends on $\Delta U_K^{-4}$ since the state is the Unruh vacuum, but unlike \eqr{TUU},
 it depends on the actual location of the two 
points through 
the $v$-dependence of 
the Jacobians $dU_K/dr$.
Using $U_K(v,r) = e^{-\kappa v}\, {\cal{U}}_K(r)$ which follows from the stationarity of the metric,
we can extract this $v$-dependence and obtain 
\be
\langle T_{rr}(v,r)\,  T_{rr}(v_0 , r_0)\rangle_{K} = 
\left( \frac{d{\cal{U}}_K}{dr}  \frac{d{\cal{U}}_K}{dr_0} \right)^2
\left( 
{4\pi} 
{\left( e^{\frac{-\kappa (v-v_0)}{2}}\, {{\cal{U}}_K} - 
e^{\frac{\kappa (v-v_0)}{2}}\, {{\cal{U}}_K^0} - i \epsilon\right)^2 }  \right)^{-2} .
\label{TTrr2}
\ee
%
There is yet another interesting way to write this correlator.
Using 
\be
1 - w^2 = 2 \kappa \left({d\ln {\cal{U}}_K \over dx}\right)^{-1},  
\label{relwU}
\ee
which follows from \eqr{rv}, we get
\be
\langle T_{rr}(v,r)\,  T_{rr}(v_0 , r_0)\rangle_{K} = 
\left(\frac{\kappa^2}{\pi} \frac{{U_K} {U^0_K}}
{(1 - w^2)(1 - w_0^2) }\,  
\frac{1}{\left( {U_K} - {U^0_K}
- i \epsilon\right)^{2}} \right)^{2} .
\label{TTrr3}
\ee

From \eqr{TTrr2}, it is clear that the correlator 
is a function of only 3 variables, $r, r_0$ and $v-v_0$, the state (Unruh vacuum) being stationary. 
Because it still
depends on 3 variables, its behavior in different 2 dimensional sections
illustrates different aspects of the correlations associated with the Hawking effect. 
When fixing  $r_0,v_0$
on ${\cal I}^+$, 
\eqr{TTrr}, function of $r, v$,
describes the correlations associated with 
a late detection on ${\cal I}^+$~\cite{Massar:1996tx}. 
Instead, at equal time 
 $v_0 = v $, 
\eqr{TTrr} 
describes the correlations in the $r,r_0$ plan
that have been accumulated in the past of that time~\cite{Iacopo08}. 

Before studying these two cases, we need to be more precise about the 
black hole 
geometries we shall work with. As explained before,
we consider profiles 
that become
constant for $r \to \pm \infty$.
To have a simple example at hand, we
choose directly ${\cal U}_K(r)$
since 
 it is the only relevant function in \eqr{TTrr2}: 
\be
\kappa \, {\cal{U}}_K(x) = - 
\left( e^{2 \kappa x} - e^{- 2 \bar \kappa x}\right) ,
\label{wex}
\ee
where $x= r - r_h$. This 
 is a kind of symmetrized version
of the Schwarzschild metric where ${\cal{U}}_K = -x \, e^{2\kappa x}$. 
Using \eqr{relwU}, one gets
\be
1 - w^2 = 
\frac{e^{2 \kappa x} - e^{- 2 \bar \kappa x}}{e^{2 \kappa x} + \frac{\bar \kappa}{\kappa} e^{- 2 \bar \kappa x}}. 
\label{relwU2}
\ee
Near the horizon, for $\kappa x \ll 1$,
one has $w^2 = 1 - 2 \kappa x $ for all values of $\bar \kappa$,  
and the asymptotic values are
$w_\infty = 0$ for $x \to \infty$, and
$w^2_{-\infty} = 1+ \bar \kappa/\kappa$ for $x \to -\infty$.

\subsubsection{Correlations to a late detection}

We fix $x_0,v_0$ on ${\cal I}^+_R$, on the future right infinity, and label it with $u_0$. We consider \eqr{TTrr}
\be
\bar T_{rr}(r,v)\vert_{ u_0, \, {\cal I}^+} \equiv
\langle T_{rr}(v,r)\,  T_{rr}(u_0,{\cal I}^+_R)\rangle_{K},
\label{Cinfty}
\ee
as 
a 1pt function. As such $\bar T_{rr}(r,v)$ is a special case of the conditional value
\be
\bar T_{rr}(r,v)\vert_{ {\Pi}_R} =  \langle T_{rr}(v,r)\,  \hat \Pi_{{\cal I}^+_R} \rangle_{K},
\label{cond}
\ee
where $\hat \Pi_{{\cal I}^+_R}$ 
is a projector that specifies 
 a state on ${\cal I}^+_R$, see App. C for a brief account, and App. C of~\cite{Brout:1995rd}
for details.
The projector can be chosen at will.
If it is taken to be $a^{as\, \dagger}_\om a^{as}_\om$
it selects the subset of states (present in the Unruh) 
which contains that asymptotic quantum
without specifying when it is detected. One can also consider the
other limit in which one only specifies the moment of detection $u_0$. In that 
case, \eqr{cond} is identical to \eqr{Cinfty}, up to an overal constant factor,
as can be seen by taking $f(\tau) = \delta(\tau - u_0)$ in Eqs.(55, 56)
of \cite{Massar:1996tx}. From this we learn the physical meaning of \eqr{Cinfty}:
it gives the mean value of $T_{rr}$ when
the $in$ 
state is Unruh vacuum, {\it and} when 
a particle is 
detected on ${\cal I}^+$ at $u=u_0$.
\begin{figure} 
\includegraphics[width=70mm]{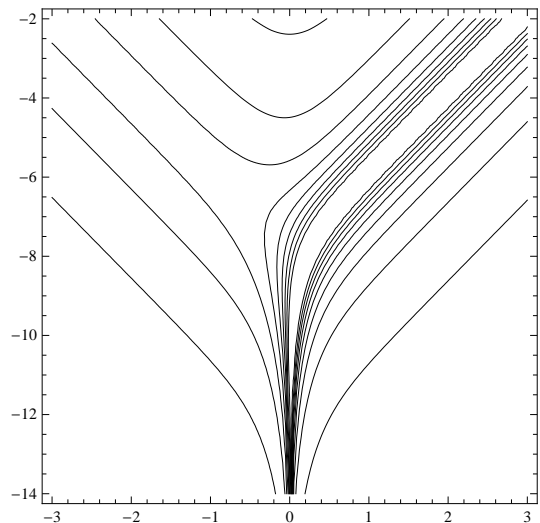}
\label{fig}
\caption{{\it The stress-energy conditional to a late detection}. We represent $\bar T_{rr}$ 
of \eqr{Cinfty} in the $x,v$ plane ($v$ being vertical) for a detection at 
$x_0 = 4,v_0=0$, outside the picture in the top right-hand region where $w(r)$ is constant. 
The horizon is at $x=0$ and 
$\kappa = \bar \kappa = 1$ in \eqr{relwU2}. 
When $\vert x \vert \geq 1$, 
on both sides, the pattern is translation invariant along the null direction because $w$ is constant. 
Instead for  $\vert x \vert < 1$, one sees the endless focusing of the null lines for $v\to -\infty$.
The 
"post-selected" partner propagates along $v + 2x = -8=v_0 - 2 x_0$, i.e. along the opposite trajectory
fixed by $x_0,v_0$. 
These features were found in \cite{Massar:1996tx} considering \eqr{cond}
for a typical Hawking quantum.}
\end{figure}

When $(r,v)$ is also on ${\cal I}^+_R$ and coordinated by $u$, \eqr{Cinfty} is given
by \eqr{Tuu} (times $4^2$), as can be seen using \eqr{TTrr3}, and $\kappa u = -\ln(-\kappa U_K)$.
In this we recover the thermal correlations~\cite{Carlitz:1986ng} of the asymptotic radiation.
When $(r,v)$ is taken on the opposite null infinity ${\cal I}^+_L$,  or 
{sufficiently far away} from the horizon so that $w$ is constant,
$\bar T_{rr}$ behaves as 
\eqr{ldTuu} when using the mirror coordinate
$\kappa \bar u =\ln(\kappa U_K)$. 
Moreover, when $(r,v)$ is near the horizon, $\kappa (r-r_h) \ll 1$,
and parameterized by $U_K$, $\bar T_{rr}$ 
behaves as in \eqr{TUU}, 
as can be seen using \eqr{TTrr} and $dU_K/dr \sim -2 $ for $r=r_h$.
In this we recover that when probed near the horizon, for 
$\kappa x \ll 1$, Unruh vacuum behaves like 
Minkowski vacuum. 

In addition to these three asymptotic behaviors, 
the non-trivial information contained in $\bar T_{rr}$  of \eqr{Cinfty} 
is the 
smooth 
interpolation from one to the other, which is represented
 in Fig. 1. From this we clearly see the gradual emergence from $v=-8$ 
of the energy flux associated with the partner on the other side of the horizon.
What is non-trivial is the following. 
In the past of that time, $\bar T_{rr}$ is essentially constant along the outgoing null lines $U_K= const$.
and, as could have been expected, behaves exactly as $\bar T_{UU}$ would in Minkowski.
On the contrary, near the horizon and in the future, $\bar T_{rr}$ behaves very differently 
since the lines $\bar T_{rr}=const.$
cross the horizon. This peculiar behavior could not have be found had we studied 
the 2pt function of \eqr{TUUC} 
because the latter obeys \eqr{TUU} and depends only on $U$, even in a black hole geometry. 
This establishes that the use of $T_{rr}$ of \eqr{Trr}
with $r$ affine 
is truly necessary. It should be also noticed that the above behavior of $\bar T_{rr}$
{cannot} be found in Minkowski using affine coordinates either. In fact, it is characteristic
 of pair creation processes, as can be seen by comparing Fig.1 to Fig. 1.1 of~\cite{Brout:1995rd}
which describes pair creation in an electric field.
As noticed in~\cite{Massar:1996tx}, these properties of  $\bar T_{rr}$ provide a 
clear answer to the long standing question: where is a Hawking quantum "born" ?~\cite{Unruh1977}.


\subsubsection{Equal time correlations}

We consider \eqr{TTrr} at equal EF time $v=v_0$. Since $U_K = e^{-\kappa v} {\cal U}_K$, 
the correlator
\be
C_K(r,r_0) = \langle T_{rr}(r)\,  T_{rr}(r_0)\rangle_{K} = \Big(\,  \partial_{r} \partial_{r_0}
\langle \phi(r)\, \phi(r_0)\rangle_{K} \, \Big)^2,
\label{Crr}
\ee 
is given by \eqr{TTrr3} with $U_K$ replaced by ${\cal U}_K$.
Using ${\cal U}_K$ of \eqr{wex}, 
$C_K(r,r_0)$ 
 diverges as $\sim(r -r_0)^{-4}$ when $r \to r_0$, 
as expected since the Unruh vacuum is a Hadamard state. 
Moreover, 
when $r\to \infty$, since ${\cal U}_K \sim - e^{-2\kappa r}$, one has
\be
C_K (r,r_0)= \frac{\kappa^4}{16 \pi^2 \sinh^4(\kappa(r-r_0))}.
\label{CWC}
\ee
The thermal noise of \eqr{Tuu} associated with the asymptotic radiation  
is properly encoded in $C_K$ 
since at fixed $v$, $du= -2 dr$.
Similarly, when $x$ and $x_0$ are on  opposite sides of the horizon,
since ${\cal U}_K \sim - e^{-2\bar \kappa \vert x \vert}$ for $x \to -\infty$,
one asymptotically gets 
\be
C_K(r,r_0) = \frac{\kappa^2 \bar \kappa^2}{16 \pi^2 \cosh^4(\bar \kappa x + \kappa x_0))},
\label{BC}\ee
thereby recovering 
\eqr{ldTuu}, and making contact with \cite{Balbinot:2007de,Iacopo08},
see Fig. 2 on the left. 
\begin{figure} 
\includegraphics[width=60mm]{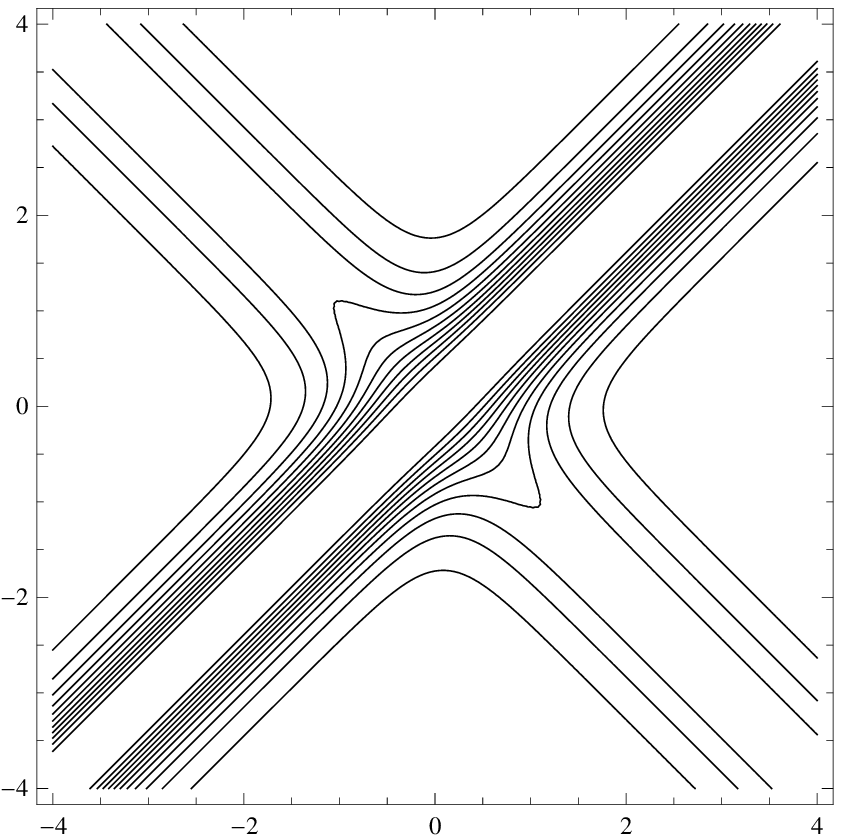}\quad \quad  \includegraphics[width=60mm]{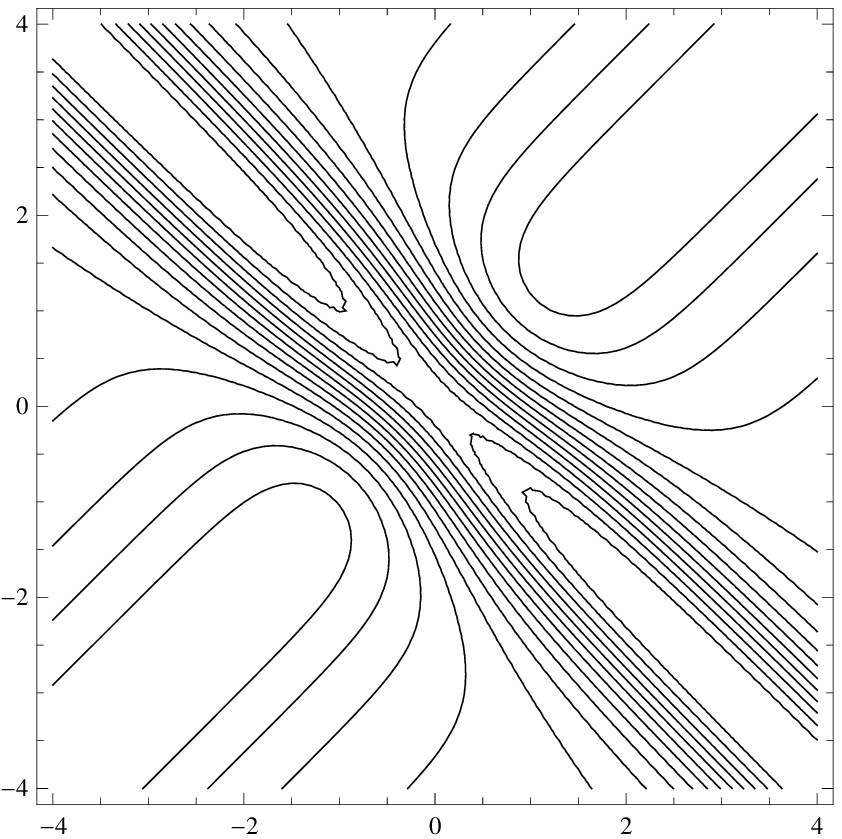}\label{fig2}
\caption{{\it Equal time correlations.} On the left, we represent 
\eqr{Crr}, and on the right, 
 \eqr{Sr}, both in the metric of Fig.~1. 
The horizon is at $x, x_0=0$. 
 On the left, the signal diverges for $x \to x_0$
whereas the pattern along $x+x_0= 0$
represents \eqr{BC}. It is 
translation invariant in $x-x_0$ 
once $w$ has reached a constant. On the right,  the subtracted 
$S_K$ 
is everywhere finite and regular. It is dominated by the 
correlations across the horizon. 
The subdominant patterns centered along $x-x_0= 0$ 
are due to the fact that the un-subtracted 
correlator in 
\eqr{Sr} 
decreases faster than the subtraction.
On the diagonal, 
$S_K(x,x) = \langle T_{rr}(x)\rangle^{\rm ren}_K$,
the renormalized flux 
 of \eqr{Tuuloc}.}
\end{figure}
To further investigate the gradual change from near horizon
configurations 
to long distance correlations,
we consider 
the subtracted 2pt function\footnote{For simplicity we worked with the square root of \eqr{Crr} 
 rather than $C_K$ itself. 
The reason is that 
the subtraction needed to obtain a finite expression for $r \to r_0$ is more complicated, as it requires three terms.
The subtracted correlator of $\langle T_{rr} T_{rr}\rangle_{K}$ possesses the same behavior as $S_K$.}
\be
S_K (r,r_0)= \partial_{r} \partial_{r_0}\left[ 
\langle \phi(r)\, \phi(r_0)\rangle_{K} - 
\frac{(-1)}{4\pi} \ln(r-r_0)\right].
\label{Sr}
\ee
The substraction term is the same as in \eqr{localG}, but considered here for all values 
of $r-r_0$.
From the the right plot in Fig. 2, we see that the {\it residual} signal is free of
UV divergences and 
describes the emergence of the pairs 
on distances $\kappa (r-r_h)= \kappa x \sim 1$, i.e. characterized by the geometry. 

One first notices that the correlations across the horizon in $S_K$
are negative (as in inflation~\cite{Campo:2003pa}). In fact the 
correlator 
$ \langle \partial_{r} \phi\,\partial_{r_0} \phi\rangle_{K} $ 
is negative "everywhere", as can be seen from \eqr{TTrr2} and \eqr{wex}. We have added quotation marks because this correlator is  a distribution.
In fact, the coincidence point limit is ruled by the $i \epsilon$, see \eqr{TUU}.
It specifies that 
(the real part of) $ \langle \partial_{r} \phi\,\partial_{r_0} \phi\rangle_{K} $ 
diverges positively for $r \to r_0$ same $v$, and it ensures that  the integral
$\int_{-\infty}^\infty dU  \langle \partial_{U}\phi\,\partial_{U_0}\phi\rangle$ 
vanishes in the Minkowski vacuum. (This is reminiscent 
of the behavior of $\langle T_{UU} \rangle_{FR}$, 
the mean 
flux evaluated in the Fulling-Rindler vacuum~\cite{Parentani:1993yz}.)
We also 
notice that for close points, $S_K$ is
positive because the subtraction is larger than the "bare" term. 
Moreover when evaluated at the same point, 
$S_K(x,x)$ is equal to $\langle T_{rr}(x)\rangle^{\rm ren}_K$
 of \eqr{Tuuloc}. For $x \to \infty$  one 
 verifies that $S_K(x,x) = (-1/3)\times S_K(x,-x)$.
%
Finally we 
notice that Fig.2. is symmetrical under  $x \to -x$. This follows
from the symmetry 
of $1- w^2$ in \eqr{relwU2} when $\kappa = \bar \kappa$.
When $\bar \kappa/\kappa \gg 1 $, this 
is lost, and $1- w^2$ resembles more to Schwarzschild.
We hope to report on this case soon.

\subsection{Time-dependent growth of flux and correlations}

So far we considered the stationary correlation patterns
found in the Unruh vacuum. 
We now consider the early transient effects. 
As already said in Sec. III.B, they 
are not universal. However
their late time behavior 
is 
when the state contains no high frequency excitations, i.e. when it is an Hadamard state.
There is a simple and efficient way to characterize this behavior.
It consists in assuming that the
initial state specified at $v=v_{\rm in}$ is the Minkowski vacuum. (This state can be found
in a gravitational 
collapse when the infalling matter is a light-like thin shell~\cite{Massar:1996tx}.)
In this state, $v= v_{\rm in}$, the 2pt function at  
 is 
$ G^{\rm in}(r,r_0) 
= \frac{-1}{4\pi} \ln(r-r_0 + i\epsilon)$, as in 
\eqr{localG}.
In 
the future, one has 
\be
G^{\rm in}(r,v; r_0, v_0) = -\frac{1}{4\pi} \ln(X^{\rm in}(r,v)-X^{\rm in}(r_0,v_0) + i\epsilon),
\label{localG2} \ee
where $X^{\rm in}(r,v)$ gives the value of $x = r- r_h$ hit 
by the outgoing null ray
issued from $r,v$ when it crosses $v_{\rm in}$. In this state, the mean flux of \eqr{Tuuloc}
becomes 
\be
\langle T_{rr}(r,v)\rangle^{\rm ren}_{\rm in}
= \frac{1}{12 \pi} 
\left( \frac{dX^{\rm in}(r,v)}{dr} \right)^{1/2}  \partial_{r}^2 
\left( \frac{dX^{\rm in}(r,v)}{dr} \right)^{-1/2},
\label{Tuuloctd}
\ee
and the correlator of \eqr{TTrr} is 
\be
\langle T_{rr}(v,r)\,  T_{rr}(v_0 , r_0)\rangle_{\rm in} = 
\left( \frac{dX^{\rm in}}{dr}  \frac{dX^{\rm in}}{dr_0} \right)^2
\left( \frac{1}{4\pi}\frac{1}{({X^{\rm in}(v,r)} - {X^{\rm in}(v_0,r_0)} + i \epsilon)^2 }  \right)^2 .
\label{TTrrtd}
\ee
In both expressions one has simply replaced $U_K$, the Kruskal coordinate encoding the Unruh vacuum,
by $X^{\rm in}$ which encodes the time-dependent state which is vacuum at $v_{\rm in}$. 
\begin{figure} 
\includegraphics[width=55mm]{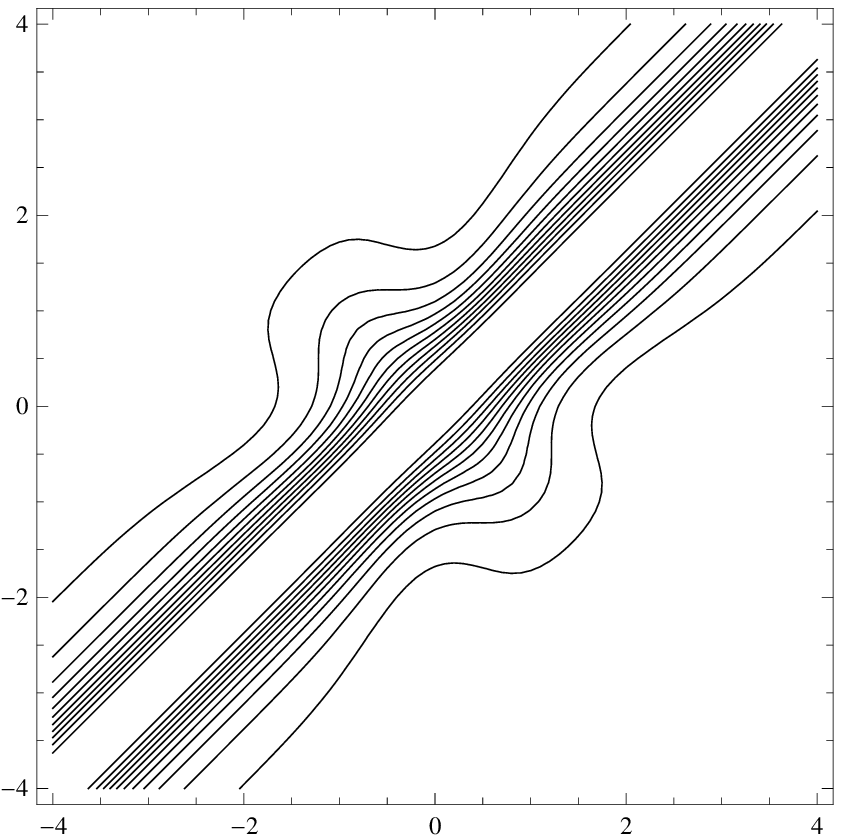}\quad \quad  \includegraphics[width=55mm]{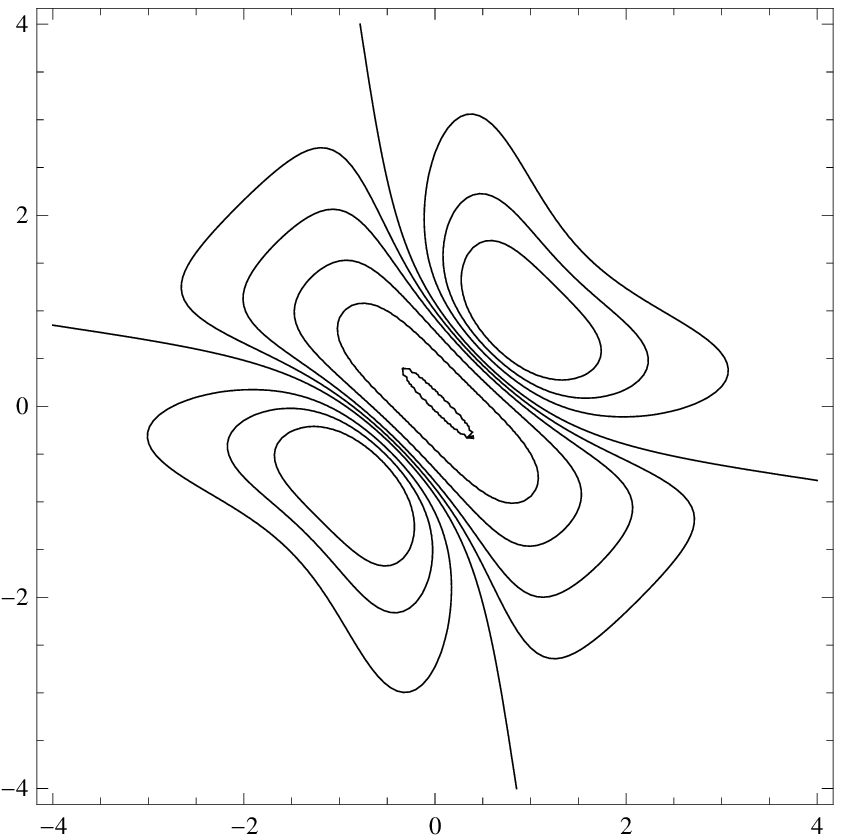}\\
\includegraphics[width=55mm]{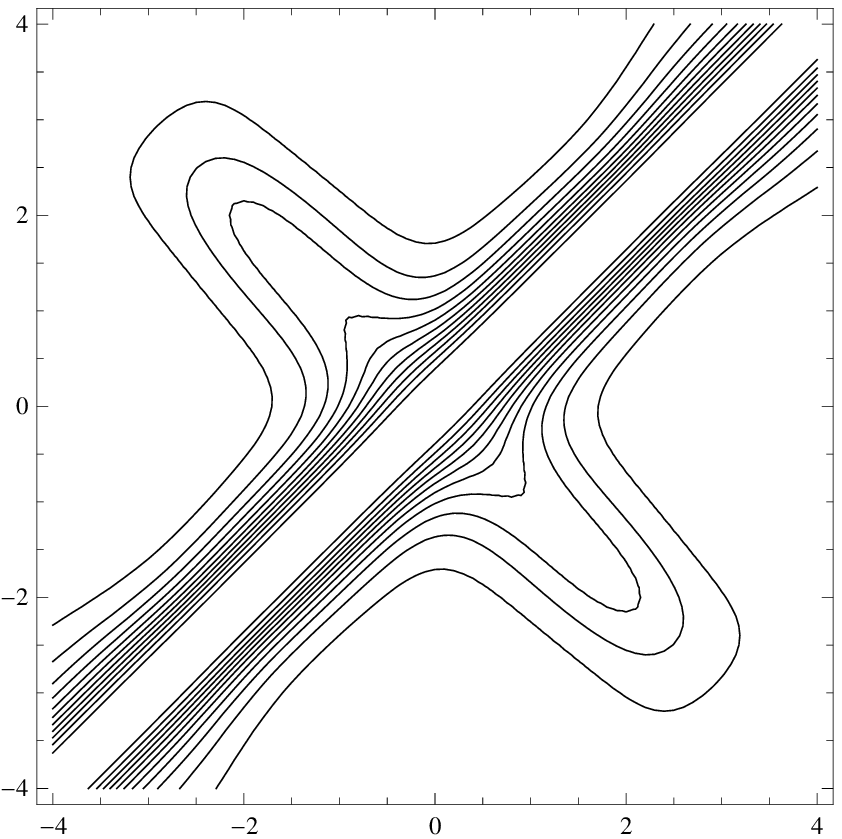}\quad \quad  \includegraphics[width=55mm]{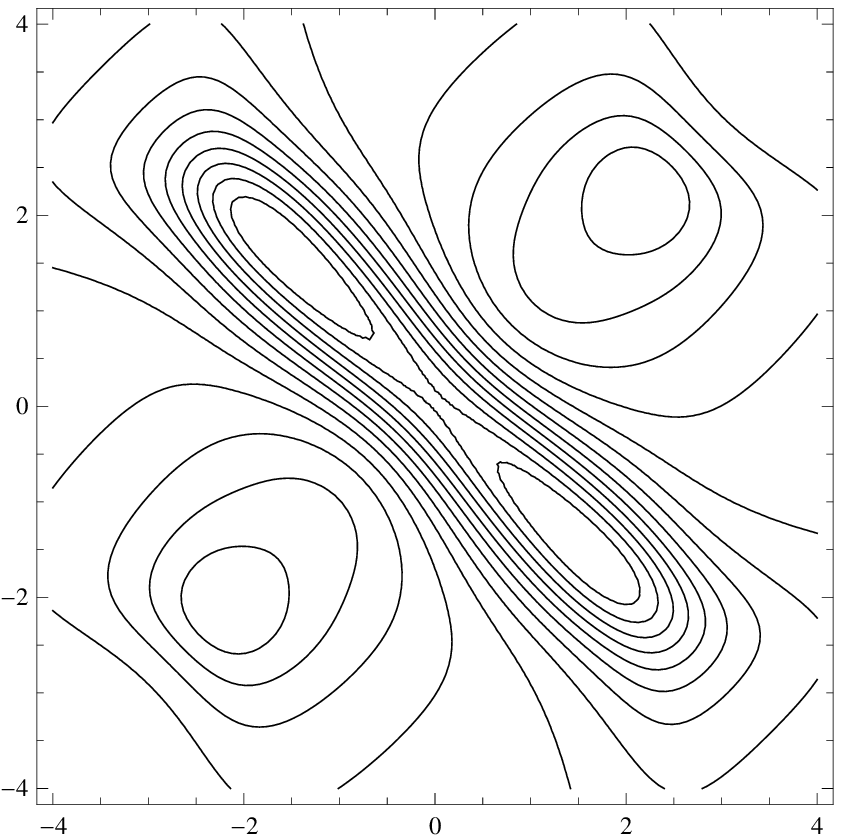}
\label{fig3}
\caption{{\it The growth of equal time correlations.} 
We represent \eqr{TTrrtd} at equal time on the left,
 and  \eqr{Sr} on the right, 
after a lapse of time $\kappa (v - v_{\rm in}) = 1$ in the upper plots, and $\kappa (v - v_{\rm in}) = 4$ in the lower ones.
On the left, 
one observes the growth of the correlations across the horizon centered along $x+x_0 = 0$.
One also observes a narrowing 
 of the correlations centered along $x = x_0 $.
On the right,  
$S_{\rm in}(r,r_0)$ displays two distinct features. A strong signal associated with the 
building up of the 
correlations across the horizon, and 
subdominant patterns on both sides of the horizon 
due to the growth 
of thermal correlations of \eqr{CWC}.
At late times, both patterns asymptote to the stationary ones of Fig. 2.}\end{figure}

In the metrics of \eqr{relwU2},
these expressions flow towards 
 \eqr{Tuuloc} and \eqr{TTrr}, exponentially  in $v-v_{\rm in}$ 
in the near horizon region
where $\partial_r w = \kappa$, 
and linearly when $\partial_r w \sim 0$. 
To show this
we 
make use of ${\cal U}_K(x)$, 
 solution of \eqr{relwU}.
Calling $X({\cal U}_K)$ the inverse function, 
and using $ U_K(x,v)= {\cal U}_K(x) e^{\kappa v}$, we obtain
\be
X^{\rm in}(r,v) = X\left[{\cal U}_K(x) \,  e^{-\kappa (v - v_{\rm in})}\right],
\ee
for all profiles $w^2(x)$.
To give a simple example, we use \eqr{wex} with $\bar \kappa = \kappa$, and we get
\ba
2 \kappa X^{\rm in}(r,v) &=& 
{\rm arcsinh}\left( e^{-\kappa (v - v_{\rm in})}\sinh(2\kappa x)\right),
\nonumber \\
\frac{dX^{\rm in}}{dx} &=& \frac{\cosh(2\kappa x)}{\left[ e^{2\kappa (v - v_{\rm in})} 
+ \sinh^2(2\kappa x)\right]^{1/2}}.
\label{exX}
\ea
In Fig. 3, on the left, we  represent 
\eqr{TTrrtd} evaluated at equal time
for different values of $v-v_{\rm in}$.
In these plots, one 
clearly sees the growth of the correlations across the horizon with a rate given by $\kappa$~\cite{Iacopo08}.
One also observes a narrowing of the spread of the dominant correlations centered along $x = x_0 $.
This is due to the 
progressive replacement of vacuum correlations $\sim 1/(x-x_0)^{2}$ by the thermal ones
$\sim \kappa^2/\sinh^{2}\kappa(x-x_0)$ of \eqr{CWC}.
On the right plots, the subtracted function $S_{\rm in}(r,r_0)$, 
the equivalent of \eqr{Sr} evaluated in the $in$ vacuum, displays 
both the growth of the long distance correlations, 
and the modifications of the local correlations.
 \begin{figure} 
\includegraphics[width=59mm]{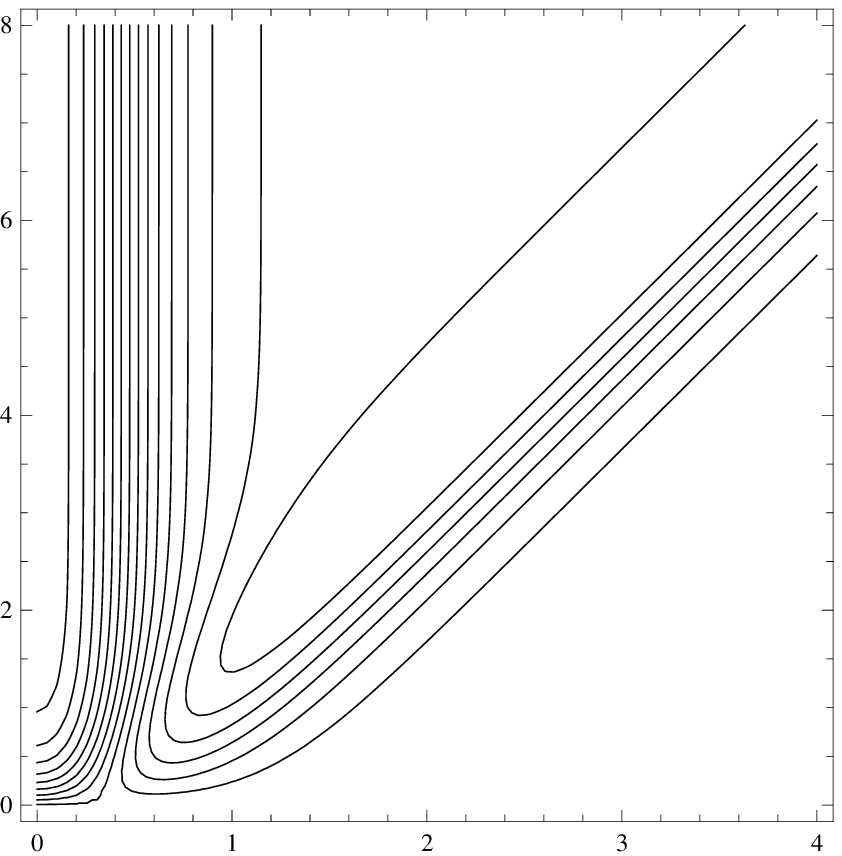}\quad \quad \includegraphics[width=60mm]{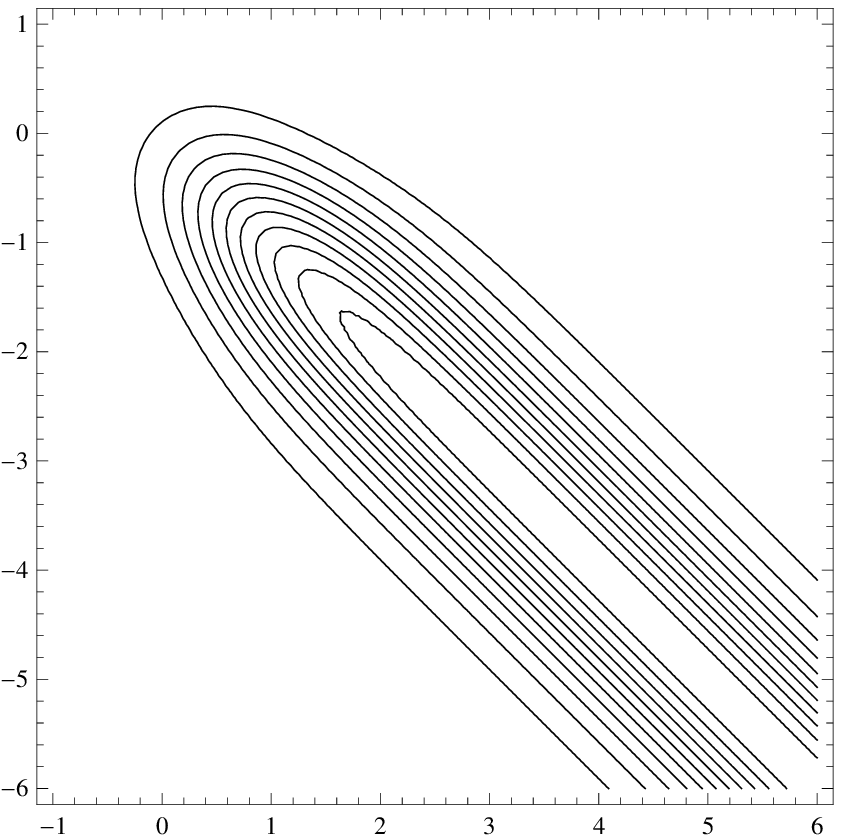}\label{fig4}
\caption{{\it Time dependence of the mean flux and of the asymptotic correlations.}
On the left, we represent \eqr{Tuuloctd} with \eqr{exX}
from the onset of the vacuum at $v=0$
till $\kappa v = 8$, and for $\kappa x$ from $0$ to $4$ (It is symmetric under $x \to -x$). 
The transients propagate on null lines $u=v-2x \sim 0$. After they passed,
 $\langle T_{rr}\rangle_{\rm in}$ is $v$-independent.
On the right plot, we represent $\bar T_{uu}(\bar u)\vert_{u_0}$ of \eqr{Cinfty} in the $u_0,\bar u$ plane
with $\bar u$ vertical, with $u_0$ and $\bar u$ defined on the future null infinity $v_0=v=\infty$. 
For $u_0< 0$, before the transients, there is no correlations across the horizon. Instead for $u_0 > 3$, $\bar T_{uu}(\bar u)\vert_{u_0}$
only depends of $u_0+\bar u$, and is given by \eqr{ldTuu}.}
\end{figure}

It is worth analyzing these time dependent effects through two other perspectives.
In Fig. 4, on the left, we present 
the mean flux of \eqr{Tuuloctd} in the $x,v$ plane from the onset of the vacuum at $v=v_{\rm in}=0$. 
$\langle T_{rr}\rangle_{\rm in}$ 
contains transients which propagate along $u=v-2x \sim 0$. 
It then reaches a constant $x$-dependent profile.
For $x\to \infty$ one recovers the standard value $\kappa^2/12 \pi$~\cite{Davies:1976ei}. 
In the metric of \eqr{wex} with $\kappa= \bar \kappa$, $\langle T_{rr}\rangle_{\rm in}$ 
 crosses $0$ for $\sinh 2x = \sqrt{2}$, i.e. $x \sim \pm .57$, and on the horizon, 
it is negative and equal to $ -2\kappa^2/12 \pi$. When including gravitational back-reaction effects,
this term participates to the evaporation of the black hole, see~\cite{Parentani:1994ij}
for a numerical analysis in similar settings.  
On the right plot, we show 
$\bar T_{\bar u \bar u}(\bar u)\vert_{u_0}$ of \eqr{Cinfty}, the 
flux correlated to a late detection at $u_0$ on ${\cal I}^+_R$, 
that we evaluate on ${\cal I}^+_L$ and parameterize with $\bar u$, the mirror coordinate $\kappa \bar u = 
\ln \kappa U_K$. There is no correlations from ${\cal I}^+_R$ to ${\cal I}^+_L$ for $u_0< - 1$ in conformity 
to the fact that no Hawking radiation as yet reached the null infinites. Then for positive $u_0$, the
correlations settle to a stationary pattern centered around $u_0 + \bar u = 0$ found in the Unruh vacuum.

In conclusion it is interesting to observe that, even though the transients
give rise to a higher value of the mean flux, they are {\it not} associated 
with stronger correlations across the horizon, and this because, unlike the steady Hawking radiation, 
the transients are not composed of entangled pairs of opposite frequency $\om$.
This can be checked 
by 
comparing 
$\bar T_{uu}(\bar u)\vert_{u}$ to 
the asymptotic flux $\langle T_{uu}\rangle_{\rm in}$ evaluated for $x,v \to \infty$,
with 
$u= v - 2x $ fixed: 
\ba
\langle T_{uu}(u)\rangle^{\rm ren}_{\rm in}
&=& \frac{\kappa^2}{48 \pi} 
\frac{1 + e^{-2 \kappa u}}{\left( 1+ e^{-2 \kappa u} /4\right)^{2}},\nonumber \\
\bar T_{\bar u \bar u}(\bar u)\vert_{u_0} 
&= &\left(\frac{\kappa^2}{\pi}\right)^2  \Big( \left({e^{2 \kappa u_0} +1/4}\right)
\left( 
{e^{-2 \kappa \bar u} +1/4} \right)
{\left( {\rm arcsinh}\left( e^{-\kappa u_0}/2 \right) +  {\rm arcsinh}\left( e^{+\kappa\bar u}/2 \right)  \right)^4}
\Big)^{-1} .
\label{TTrrtd2}
\ea

\section{Dispersive theories and Analogue black holes}


In a non-homogeneous medium, linear density fluctuations
obey a relativistic d'Alembert equation in a curved 
metric 
when their wave lengths are  larger than the inter-atomic distance~\cite{Unruh:1980cg}. 
Instead, for shorter wave lengths, the propagation becomes dispersive~\cite{Jacobson:1991gr}.
Assuming the speed of sound is constant and set to $1$, it can be 
described by
\be
\Om^2 = (\om - w p)^2 = F^2(p^2),
\label{dr}
\ee
where  $\Om$ is the co-moving frequency measured w.r.t. to
the atoms, $w$ the velocity of the fluid, and $p,\om$ the wave vector 
and the frequency measured in the lab.
The dispersion can be incorporated in
a modified 
field equation~\cite{Unruh:1994je}  \be
\left(\partial_\tau 
+ \partial_x w \right) \left( \partial_\tau 
+ w \partial_x \right) \phi - \partial^2_x \phi \pm \frac{1}{\Lambda^2} \partial_x^4 \phi =0 ,
\label{modeeq}
\ee
which reduces to the relativistic one when sending the dispersive scale $\Lambda \to \infty$.
For simplicity, we chose  
quartic, super (+) or subluminal (-) dispersions: $F^2 = p^2 \pm p^4/\Lambda^2$.

\eqr{modeeq} can then be used to study the impact of dispersion on Hawking radiation.
In ~\cite{Unruh:1994je}, the thermicity and the stationarity of the asymptotic 
radiation have been shown to be robust,
i.e. hardly affected by dispersion when  $\kappa \ll \Lambda$. 
This is 
sufficient for recovering  
\eqr{CWC}.
By constructing 
wave packets of $in$ modes, 
it was then shown~\cite{Brout:1995wp} that 
at large distance from the horizon one also recovers the 
correlations between Hawking quanta and their partners, and this, even though
the early propagation was radically affected by dispersion.
This second aspect 
is sufficient for obtaining 
\eqr{BC}, Fig. 2, 
and the late time properties of 
Fig. 1. 

To clarify this, 
we shall compare 
the properties
of wave packets of $\phi^{in}_\om$, the $in$ modes of \eqr{modeeq},
\be
\bar \phi_{\bar \om}(\tau, x) = \int d\om  \,   e^{- i \om \tau} \phi^{in}_\om(x) \, \bar f_\om,
\label{wpbhx}
\ee
where $\bar f_\om$ 
 selects the wave packet,
with the 
correlation function
\ba
G^{in}(\tau, x ;\tau_0,x_0)& \equiv&  \langle \phi(\tau, x) \, \phi(\tau_0,x_0) \rangle_{in} \nonumber 
\\
&=& \int d\om   \, 
e^{- i \om \tau} \phi^{in}_\om(x)\,  [ e^{- i \om \tau_0} \phi^{in}_\om(x_0)]^* ,
\label{Gbhx}
\ea
evaluated in the stationary $in$ vacuum.
When dispersion is weak, the similarity of 
expressions
guarantees that  
similar patterns will be found. However, 
since dispersion grows as approaching the horizon, the way one probes the correlations,
i.e. by extracting some limited range of 
 $\om$ through $\bar f_\om$ in \eqr{wpbhx},
or not as in \eqr{Gbhx}, can lead to different behaviors. 
We now review the relevant 
points for achieving this comparison following~\cite{Brout:1995wp,Jacobson:2007jx};
other treatments are mentioned in~\cite{Jacobson:1999zk}.


\subsection{Kinematics}

From a relativistic point of view, 
the presence of dispersion 
defines a preferred frame~\cite{Jacobson:1996zs} which allows to define new scalars. 
This is best seen by "covariantizing" \eqr{dr},  i.e. 
by introducing a unit time-like vector field  $u^\mu$,
and viewing the field $\phi$ 
as propagating on a manifold endowed with both the metric and $u^\mu$.
Then the energy in the preferred frame,
and the spatial momentum perpendicular to it are respectively
\be
\Om = u^\mu p_\mu, \quad p = s^\mu p_\mu,
\label{Om2}
\ee
where $- u^2 = 1 = s^2$ and $s^\mu u_\mu= 0$. The PG coordinates used in \eqr{modeeq}
can then be {invariantly} 
defined by $\partial_x = s^\mu \partial_\mu$ and $d\tau = u_\mu dx^u$.
The field $u^\mu$ also defines the scalar density $\rho = u^\mu u^\nu T_{\mu\nu}$
which corresponds to the proper energy that observers following
$dx^\mu/d \tau = u^\mu$ would measure. 
It is interesting to note that in the hydrodynamical limit one gets
\be
\rho =  T_{xx} = (\partial_x \phi)^2.
\label{rho}
\ee
 In other words, 
$\rho$ 
coincides with $T_{rr}$ of \eqr{Trr} with
 $r$ defined by \eqr{rv}. This non-trivial correspondence follows from the affinity of $r$ at fixed 
$v$ and $\tau$,  
see the remarks after \eqr{EF}.

In the same spirit, we notice that the surface gravity measured with respect to the preferred frame is also 
scalar. It is given by the expansion~\cite{Jacobson:2008cx} 
$\theta = u^\mu_{;\mu} = \partial_x w$ evaluated on the horizon.
(The second expression is valid in PG coordinates).
The ambiguity of the scale of the surface gravity in covariant theories is thus removed 
when using 
$u^\mu$.

\subsection{The modified modes}


In linear field theories, the modifications of the stress energy correlations due to 
dispersion will stem from the modifications of the modes, solutions of \eqr{modeeq}. When $\Lambda \gg\kappa$,
 these  
are modifications localized near the horizon, for $\kappa(r - r_h)= \kappa x \ll 1$. 
As noticed in \cite{Brout:1995wp} it is appropriate to work in the $p$-representation (with $x = i \partial_p$)
with $w(x)$ linearized: $w= -1 + \kappa x  = -1 + i \kappa \partial_p$.
 In this 
representation, 
\eqr{modeeq} becomes 
\be
 \left(\omega - p w \right) \left( \omega -  w p \right)\tilde{\phi}_{\omega} = 
 F^2\,  \tilde{\phi}_{\omega}. 
\ee
Using $w= -1 + i \kappa \partial_p$,
the modified modes 
have the form, for details see the Appendix of~\cite{Jacobson:2007jx},
\be
\tilde{\phi}_{\omega} = \tilde{\phi}^0_{\omega} \times e^{- i p/\kappa } \chi(p),
\ee
where $\tilde \phi^0_\om = 
\vert p\vert ^{-i \om/\kappa - 1}$
is the standard 
dispersion less mode, 
and where $\chi$ obeys
\be
 - \kappa^2 \partial_p^2 \,  \chi = \frac{F^2}{p^2}\, \chi= H^2\, \chi .
\label{ceq}
\ee
 The $in$ state 
which generalizes the notion of the Unruh vacuum, the "free falling vacuum",
is characterized by the positive norm modes which contain only positive $\Om$ of \eqr{Om2}. 
These 
$in$ modes 
are related to the Unruh modes, see \eqr{Ump}, 
\be
\tilde \phi^U_\om = \theta(p) \, \frac{ p^{-i \om/\kappa - 1}}{(4 \pi \kappa)^{1/2}},
\label{bhUm}
\ee
by~\cite{Brout:1995wp} 
\be
\tilde{\phi}^{in}_{\omega} = \tilde{\phi}^U_{\omega} \,\times\,  e^{- i p/\kappa }  \chi(p),
\label{bmpsm}
\ee
where $\chi(p)$ is the solution of \eqr{ceq} with a 
Wronskian equal to $\chi^* \partial_p \chi -\chi \partial_p \chi^* = 2i /\kappa$.
The corresponding WKB solution is
\be
\chi(p) = \frac{1}{(H)^{1/2}} \, 
 \exp\left({ i \int^p_{p_0} H(p')dp'/\kappa}\right) .
\label{Wchi}
\ee
It provides a reliable approximation 
when $\Lambda\gg \kappa $.
In the limit $\Lambda \to \infty$ fixed $p$, $H \to 1$ and $e^{- i p/\kappa }\chi \to 1$,
thereby implying that $\tilde{\phi}^{in}_{\omega}$ 
smoothly gives back $\tilde \phi^U_\om$.\footnote{At fixed $x$ 
instead, the limit $\Lambda \to \infty$ 
can be singular as some roots $p(x,\om)$ of \eqr{dr} are sent to infinity.
The $p$-WKB approximation should not be confused with the usual one defined in $x$-space. 
For the Airy function, 
the modes in $p$ are exactly given by their $p$-WKB approximation.
Similarly here, the corrections to 
\eqr{Wchi}
are negligible when $\Lambda/\kappa \gg 1$. 
This has been confirmed by numerical analysis,
see~\cite{Macher:2009tw} for a detailed study. 
From now on we neglect them and work in the adiabatic approximation with \eqr{Wchi}.}

To understand the impact of dispersion, 
we now study the 
characteristics of 
\eqr{modeeq} 
since the maximum of correlations will be localized along them. 
Having already the 
modes in $p$-space, 
the simplest way to get them 
is to consider 
\eqr{wpbhx}  in $p$-space, and
 look for the stationary phase condition in $\om$.
Using \eqr{bmpsm}, since $\chi$ is independent of $\om$,
 one gets
\be
p(\tau) = p_0 \, e^{- \kappa \tau},
\label{pevol}
\ee
%
irrespectively of the dispersion relation $F$, and thus as in relativistic theories.

To get the modified characteristics in $x$, we use 
 $\om - wp = F$, the root of \eqr{dr}
describing the right moving modes.
Using
$w\sim -1 + \kappa x$,
 one gets~\cite{Balbinot:2006ua}
\ba
\frac{\kappa}{2} \left[ x_\om(p) - x_{-\om}(p)\right] &=& \frac{ \om} {p} , 
\nonumber \\
\frac{\kappa}{2} \left[ x_\om(p) + x_{-\om}(p)\right]
 &=& 1- H(p). 
\label{modchar}
\ea
The first equation is again 
independent of $F$ and coincides what is found in 
relativistic theories, namely,
when propagated backwards in time, 
pairs of characteristics pill up 
exponentially in PG time.
The second equation tells us that the "center of mass" of a pair which is centered on the relativistic
horizon for $1-H \ll 1$, i.e. $p \ll \Lambda$, gradually moves 
away 
as $p$ increases. 
For sub (super) luminal  
dispersion, $H< 1$ ($H> 1$), the pair is sent outwards (inwards). 
For quartic dispersion 
the momentum at the turning point is given by
$p_{t.p.}^3 = 2 \Lambda^2 \vert \om\vert $~\cite{Corley}.
Because of this movement away from the horizon, 
the increase of $p$ and the focusing of  
$x_\om - x_{-\om}$ will stop when the pair reaches $\vert \kappa x \vert \sim D$,
where 
the gradient $\partial_x w$ drops down. 
A straightforward calculation gives that the focusing stops for $p\sim \Lambda D^{1/2}$. 
 
The end of the focusing and the movement away from the horizon are the 
principal consequences of dispersion. 
They imply that the early properties 
of Fig. 1 are inevitably modified, as we shall see below. 

\subsection{Correlations from wave packets}

To show how dispersion affects 
the correlation pattern encoded in \eqr{wpbhx}, 
we need 
\eqr{bmpsm} in $x$-space
\be
\phi^{in}_\om(x) = \int_0^\infty {dp \over (2\pi)^{1/2} }  e^{ipx} \tilde\phi^{in}_\om(p).
\label{phi_in}
\ee
Far away from the turning point, one can 
evaluate this integral at the saddle point approximation (since it is reliable~\cite{Brout:1995wp,Corley}),  
and decompose 
\eqr{phi_in}  in terms of
outgoing modes defined for low momenta. Doing so one finds
\be
\phi^{in}_\om(x) = \alpha_\om \left[ \theta(x) \, \varphi_\om(x) + z_\om \, \theta(-x) \,(\varphi_{-\om}(x))^* \right],
\label{aBog}\ee
where $\varphi_\om(x)$ is the $x$-WKB mode of \eqr{modeeq} with unit norm.
Explicitely it is given by
\be
\varphi_\om(x) =  \sqrt{\frac{\partial p_\om(x)}{\partial \om}}\,  
\frac{\exp \left(i \int^x_{x^0} dx' p_\om(x')\right)}{\sqrt{4 \pi  \Omega(p_\om(x))}},
\label{WKBmodes}
\ee
where $p_\om(x)$ is the corresponding low momentum root of \eqr{dr}.
One easily verifies that these $out$ modes are identical to the relativistic ones for $p_\om \ll \Lambda$. 
Moreover, for $\kappa \ll \Lambda$, up to a phase, one finds $z_\om = e^{- \pi \om /\kappa}$.
Hence \eqr{aBog} gives  the equivalent of \eqr{Um}.~\footnote{This approximation
is valid provided $\om$ is sufficiently small. For quartic dispersion
there is 
a critical frequency $\om_{\rm max}$, 
related to both $\Lambda$ and the
asymptotic velocities $w(\pm \infty)$, 
above which $z_\om$ identically vanishes~\cite{Macher:2009tw}. 
 From the numerical results of that ref., a good fit 
is 
$\vert z_\om^F \vert  =  e^{- \pi \om /\kappa}\, 
\left(1 - \om/\om_{\rm max}\right)^{1/4} $.\label{Macherz}}

In addition to the above two low momentum modes, there is a third saddle 
--on the left (right) of the horizon for super (sub) luminal dispersion,
in conformity with \eqr{modchar}--
which gives a 
high momentum mode.
Its WKB wave is also given by \eqr{WKBmodes}
with $p_\om$ being 
 the unique large positive real root of \eqr{dr}. One 
verifies that its overall coefficient 
is unity in conformity with the fact that it describes the incoming
 mode that shall be scattered.
%

Therefore, considering 
\eqr{wpbhx} 
with $\bar f_\om$ 
centered 
around 
$0 < \bar \om \ll \om_{\rm max} $, we get two results. 
First, at late times, using \eqr{aBog} and \eqr{WKBmodes},  
one finds two low momentum packets following \eqr{modchar} with $\om= \pm \bar \om$, where the negative
frequency packet 
has its amplitude reduced by $z_{\bar \om}$, as in \eqr{wp}.
Since $z_\om = e^{- \pi \om /\kappa}$, and since both $\varphi_\om$ of \eqr{aBog} 
behave as relativistic modes once $p$ is small enough ($F-p \ll p$), at large distances, 
the pattern is {\it indistinguishable} 
from 
the relativistic one obtained by replacing \eqr{bmpsm} by \eqr{bhUm}.

Second, at early times, only the incoming high momentum mode 
constructively interferes. It has a mean positive frequency $\bar \om$,
 follows the second line of \eqr{modchar} 
with $p \gg \bar \om$, and leaves the near horizon 
region with $p \sim \Lambda$.
This is completely different from what is obtained in the relativistic case. 
Indeed, 
using \eqr{bhUm}, 
$p$ would keep increasing  for ever following \eqr{pevol},
and the spread in $x$ correspondingly decrease as $\sim 1/p$. 
%

From the analysis of wave packets, we have thus reached two important results. 
On one hand, the low momentum (late time) properties of the relativistic pattern of Fig. 1
is unaffected by dispersion. On the other hand, the early properties of this pattern will
be radically affected by dispersion since the peak of correlations will follow  \eqr{modchar},
as represented in Fig. 4 of~\cite{Brout:1995wp}. 
To further investigate how dispersion affects the properties of Figs. 1 and 2, we now consider
the pattern encoded in \eqr{Gbhx} rather than in \eqr{wpbhx}.


\subsection{Correlations in energy density}


We start with 
the correlations of a relativistic field expressed in the present language.
Since $\Om = p$, 
the correlation function of $\rho$ of \eqr{rho} is 
\be
\langle\rho(x,\tau) \rho(x_0,\tau_0)\rangle_{in} =  \left( 
\partial_x \partial_{x_0}\int_{-\infty}^{\infty} d\om e^{-i \om (\tau - \tau_0)} G^{in}_\om(x,x_0)\right)^2, 
\label{prho}
\ee
where $G^{in}_\om = \phi^{in}_\om(x) (\phi^{in}_\om(x_0))^*$ is the $\om$ component of 
$G^{in}$ 
of \eqr{Gbhx}.
In the $p$-representation, using 
\eqr{bhUm}, one gets
\ba
\tilde G_K(p,\tau;p_0,0) &=& \int_{-\infty}^{\infty} d\om e^{-i \om \tau} \tilde G^{in}_\om(p,p_0)
\nonumber \\
&=& \theta(p) \theta(p_0)\,  {1 \over 2 p p_0} \, \delta(\kappa \tau + \ln(p/p_0)).
\label{Gnodisp_p}
\ea
On one hand, we recover the classical evolution law of \eqr{pevol}. On the other we learn that
in the Unruh-vacuum, at $\tau = \tau_0$, only configurations with equal values of $p$ contribute. 
%
%
There is no spread in $p$ in this state.
When inverse Fourier transform, one gets
\be
G_K(x,\tau;x_0,0) = - \frac{1}{ 4 \pi} \ln(x  - x_0\, e^{\kappa \tau}+ i \epsilon).
\label{Gnodisp_x}
\ee
We recover 
the standard result, \eqr{2ptphi}, 
expressed in PG coordinates.
At equal PG time, we notice also that 
the argument of the log is $x-x_0$. 
We shall return to this point below.
When computing $\partial_x \partial_{x_0} G_K$ 
one obtains
\be
\partial_x \partial_{x_0} G_K(x,\tau; x_0,0) = - \frac{1}{ 4 \pi} 
\frac{e^{\kappa \tau}}{(x - x_0\, e^{\kappa \tau} + i \epsilon)^2},
\label{ddcf}
\ee
which is the square root of \eqr{TTrr2} in the near horizon region where ${\cal U}_K \sim -x$.


When introducing dispersion, 
$\rho$
receives corrections with respect to $(\partial_x \phi)^2$ due to the non-linearities of $F^2$. 
This is hardly relevant for us, because in the near horizon region, the momenta $p$ are much smaller 
than the UV scale $\Lambda$. They are of course modifications in the UV sector of the theory,
but these ultra local effects are the same as in Minkowski space.
Therefore the main modifications will come from the replacement of the Unruh modes
by the modified ones.
 Using \eqr{bmpsm}, \eqr{Gnodisp_p} is replaced by~\footnote{We proceed as in \cite{Schutzhold:2010ig}. 
Nevertheless the forthcoming equations differ in several respects.} 
\ba
\tilde G^{in}(p, \tau; p_0, 0) &=& 
\tilde G_K(p, \tau; p_0, 0) \times 
\frac{\exp  i \int^p_{p_0} [H(p')- 1] dp'/\kappa}{(H(p) H(p_0))^{1/2}} .
\label{Gdisp_p}
\ea
It should be noticed that 
the frequency $\om_{\rm max}$ mentioned in footnote \ref{Macherz}, 
will cut out the integral in \eqr{prho}, below $-\om_{\rm max}$ for superluminal dispersion,
and above $\om_{\rm max}$ for sub luminal.
We ignored for this UV cutoff in computing \eqr{Gdisp_p} because \eqr{bmpsm} is no longer trustworthy
anyway when $\om \to \om_{\rm max}$. In fact, in the adiabatic approximation
of \eqr{Wchi}, there are cancelling
errors, in that the next equation can be shown to be exact. 


\subsubsection{Equal time correlations}

At equal times, \eqr{Gdisp_p} gives
\ba
\tilde G^{in}(p,p_0, \delta \tau= 0) &=&  \theta(p) \, \frac{\delta(p -p_0)}{2 \Om(p)} ,
\label{etGdisp_p}
\ea
and in the $x$-representation, 
one has 
\ba
G^{in}(x,x_0, \delta \tau= 0) &=&  \int_0^\infty {dp \over 4\pi } {e^{ip(x-x_0) }\over \Om(p)}.
\label{Geqt}
\ea
The only effect of dispersion is to replace in the denominator
the relativistic law $\Om = p$ by $\Om = F(p)$. 
Therefore \eqr{etGdisp_p} is {exactly} what one obtains in Minkowski 
vacuum in the preferred frame. 
The reason 
is again that $x$ is affine at fixed $\tau$: $ds^2 = dx^2$.
In fact, as in the relativistic case, see Sec. III.F, a non-stationary vacuum can be defined
at a given time but for {all} values of $x$
by 
plane waves $e^{ipx}$ with $p>0$. 
Then, 
the negligible character of the non-adiabatic corrections to \eqr{Wchi} in {the near horizon region} 
gives
 \eqr{Gdisp_p} which implies that in {\it that} region but at {\it all} times  
the vacuum 
stays characterized by $p > 0$.\footnote{
\eqr{etGdisp_p} is also 
 obtained in the adiabatic approximation, 
in cosmological backgrounds when the preferred frame is aligned along the 
cosmic frame. Moreover, 
this correspondence becomes exact ({\it beyond} the adiabatic approximation)
 when considering de Sitter space when 
the Hubble parameter $H=\kappa$ 
since 
the linearized expression $w = -1 + \kappa x$
describes this space in PG coordinates when $-\infty < x < \infty$.} 
This also implies that 
the dispersive version of the time-dependent 
\eqr{localG2}
will evolve towards the stationary \eqr{Gbhx}, as 
\eqr{localG2} evolved into 
the stationary 
 function in 
the Unruh vacuum.


From \eqr{Geqt} several consequences can be drawn.
If one probes the $in$ state for 
$\kappa \vert x \vert \ll 1$, 
the deviations w.r.t. to the relativistic case for $x-x_0 < 1/\Lambda$ ($p > \Lambda$) 
are the same as in Minkowski,
and are therefore insensitive to 
presence of the black hole.
If one probes 
 the $in$ state 
further away from the horizon and for momenta $p < \Lambda$, since $\Om \sim p$, 
\eqr{Gbhx}
will behave as the  
relativistic function, as it obeys the same equation, and possesses the same initial conditions. 
Hence the whole analysis of Sec. III.E.2 applies. 
In particular, 
as soon as $w$ is constant, $\partial_x w \ll \kappa$, 
\eqr{prho} 
will obey 
\eqr{CWC} when both points are on the same side of the horizon, 
and 
\eqr{BC} when one is on either side.~\footnote{In this we 
recover 
what has been found 
in Bose Einstein 
condensates when looking at the density-density correlation function~\cite{Balbinot:2007de,Iacopo08}.
In that case, in the hydrodynamical limit, 
the atom density fluctuation 
is given by $\partial_x \phi$, and 
the correlation corresponds to 
\eqr{ddcf}.}
%
Thus the properties of Fig. 2 (left) are not affected by dispersion when $\kappa \ll \om_{\rm max}$.
Those of Fig. 2 (right) are not either when subtracting the dispersive expression that replaces the
log in \eqr{Sr}, because $S_K$ varies on scales $~1/\kappa \gg 1/\Lambda$. 
  
The insensitivity of Fig.2 against introducing dispersion 
is quite surprising since, as discussed before, we expect that the
properties of Fig. 1 be affected by 
the drift of \eqr{modchar} 
which occurs for rather low momenta $\sim \Lambda^{2/3} \kappa^{1/3} \ll \Lambda$.
 The reason of the disappearance of the drift (at equal PG time)
is the following. In \eqr{prho},
because we are summing over $\om$, we erase the coherence in $x$-space 
that exists in each $\om$ sector,
thereby 
 recovering the 
translation invariance of the $in$ state, as in \eqr{2ptphi}.
In other words, it is only when isolating some $\om$ content out of all vacuum 
configurations that the {\it early} pattern characteristic of wave packets 
emerges. This deserves further comments.

Given \eqr{Geqt}, what can be said about the entanglement entropy ?
The 
regular behavior of the dispersive $in$ modes and the
entanglement in Fock space between states of opposite $\om$, see \eqr{sqs},
were exploited in \cite{Jacobson:2007jx} 
to argue that the entanglement entropy of a black hole is finite (in 1+1 dimensions).
However, using \eqr{Geqt} one would conclude that upon tracing over inside configurations $x< 0$,
one would obtain the same (diverging) result as in Minkowski~\cite{Bombelli:1986rw}. 
This 
conflictual result
 indicates
 that 
there is probably no unique notion of the entanglement entropy. 
Therefore to get a well defined result, it is needed to specify what one exactly means by
"tracing over the inside configurations".

We saw that the 2pt 
correlation function (at equal time) 
does not display the characteristic pattern of wave packets with a given frequency content. 
This is quite general. 
It was discussed in \cite{Obadia:2002qe} 
when studying the correlations amongst particles emitted by accelerator mirrors, 
and in a inflationary context 
in \cite{Campo:2003pa}.\footnote{These remarks raise the question of the choice of (the set of) observables
used to probe a  quantum state. To give a concrete ex.: 
in inflationary cosmology, it is generally assumed that the
large amplification experienced by primordial fluctuations erase all
quantum properties and would give a state indistinguishable from a stochastic ensemble
of classical fluctuations. In \cite{Campo:2005sv} it was shown that irrespectively of the amplification 
there exist observables exhibiting violations of Bell inequalities (for linearized modes).}
%

\subsubsection{Correlations at different times}

When $\tau \neq 0$, in the relativistic case one gets \eqr{Gnodisp_x}.
Instead, \eqr{Gdisp_p} gives
\ba
G^{in}(x,\tau; x_0, 0) &=&  \int_0^\infty {dp \over 4\pi } e^{i 
p \delta 
}
{e^{\kappa \tau/2}
\over  {[\Om(p)\,  \Om(p e^{ \kappa \tau} ) ]^{1/2}}}
\, \, \exp  i \int^{p}_{p e^{ \kappa \tau} } [H(p')- 1] dp'/\kappa, 
\label{mG}
\ea
where 
$\delta = x 
- x_0 e^{ \kappa \tau}$. 
%
The non-trivial modifications of the correlations due to dispersion are best seen 
by evaluating this integral at the saddle point approximation. The value of the saddle $p_*$ 
answers the classical question: given that one starts at $x_0, \tau_0=0$
and ends at $x, \tau$, what is the momentum at that time ?
It is given by 
\be
\kappa \delta = 
\left[ 
e^{\kappa \tau} (H(p_* e^{\kappa \tau} ) - 1 )
- (H(p_*) - 1) 
\right] .
\label{Gnoteqt}
\ee
To understand the implications of this expression, we consider three regimes.
First, if $\kappa \delta\ll 1$ and $\kappa \tau \sim 1$, the two points are almost connected by 
a 
null ray and the red-shifting effect is moderate. 
In this case $p_*/ \Lambda \ll 1$ and one recovers the relativistic behavior of \eqr{Gnodisp_x}.

Second, we study the non-trivial correlation far away from the light cone (but still in the near horizon region). 
For definiteness we restrict attention to
quartic laws $F^2 = p^2 \pm p^4/\Lambda^2$.
We expand 
\eqr{Gnoteqt}
to first order in $1/\Lambda^2$, and using $H - 1 \sim \pm p^2/2 \Lambda^2$, 
we get 
\be
\kappa \delta = \pm {p_*^2 \over 2 \Lambda^2} \, (e^{3\kappa \tau} - 1 )
\label{kd}
\ee
When  
$x = x_0 > 0$, 
irrespectively of the sign of $\tau$,
there is no (real) saddle for the $+$ sign, i.e., superluminal dispersion,
in agreement with \eqr{modchar} which says that both 
partners 
are dragged inside the black hole horizon. Instead, for subluminal dispersion, since they are both 
dragged outside, there must be a non trivial solution. 
To confirm this, 
we take 
$\kappa \tau $  such that $e^{\kappa \tau} 
\gg e^{-\kappa \tau}$. In this regime, \eqr{kd} reduces to $\kappa x_0   = p_*^2 e^{2\kappa \tau} /2\Lambda^2 $.
From this we can deduce $\om_*$ the mean value of the frequency corresponding to the trajectory
that goes from 
$x_0$ back to it in a lapse equal to $\tau$. 
It is approximatively given by
\be
\om_* = \sqrt{2}\Lambda \, (\kappa x_0)^{3/2} \,  e^{-\kappa \tau} . 
\ee
This result can also be derived using \eqr{modchar} (and applied to superluminal dispersion for $x< 0$).
Thus, when studying $G(x,\tau;x_0,0)$ at sufficiently large $\kappa \tau$,  
unlike what we found in \eqr{Geqt}, 
the correlations are now in agreement 
with the locus of constructive interferences of wave packets 
because only a limited range of frequencies centered about $\om_*$
significantly contributes. This confirms that near horizon behavior of 
\eqr{prho}
 will completely differ from that of Fig. 1, and will be similar to those of Fig. 1. of \cite{Jacobson:2007jx}.
What remains to be clarified concerns 
the profile of \eqr{prho} at early times. Namely, at fixed $x_0,\tau_0$, 
what is the trajectory 
of the maximum of \eqr{prho}, and what is its spread in $x$ as a function
of $ x_0, \tau-\tau_0$, and $ \Lambda$ ? We conjecture that both of these quantities are
ruled by $\om_{\rm max}$ of footnote \ref{Macherz}.

What can be studied~\cite{Schutzhold:2010ig} is the "off-shell"
limit of very large blue-shift $e^{\kappa \tau}\gg 1$
encoded in a backward propagation at fixed $x$ and fixed $\delta = x - x_0 e^{\kappa \tau}$.
This limit displays how dispersion tames the "trans-Planckian" behavior 
found for the relativistic field. 
In that case, \eqr{ddcf} gives
\be
\partial_x \partial_{x_0} G_K(x,0 ; x_0, -\tau) = - \frac{1}{ 4 \pi} 
\frac{e^{\kappa \tau}}{(\delta+ i \epsilon)^2},
\label{ddcfd}
\ee
how ever large is $\kappa \tau > 0$, in agreement with Fig. 1. 
In the dispersive case, using \eqr{mG} and $H(p) - 1 \sim \pm p^2/2 \Lambda^2$, 
one has 
\ba
\partial_x \partial_{x_0} G^{in}(x,0 ; x_0, -\tau) &=& 
\int_0^\infty {dp \over 4\pi } 
{p^2  e^{3 \kappa \tau/2}
\over  {[\Om(p)\,  \Om(p e^{ \kappa \tau} ) ]^{1/2}}}
\, \exp i \left(  p \delta \mp {p^3  e^{3\kappa \tau} \over 6 \Lambda^2 \kappa}\right).
\label{mGd}
\ea
When the blue shift is moderate, i.e.  $e^{\kappa \tau} < \kappa \delta \,  (\Lambda/\kappa)^{2/3}$,
\eqr{mGd} behaves as \eqr{ddcfd} plus corrections in $e^{3\kappa \tau} /\Lambda^2 \kappa \delta^3 \ll 1 $  
that can be computed perturbatively, as can be seen by changing variable $p \to q = p\delta$. 
Instead, when the blue shift is 
large: $e^{\kappa \tau} >  \kappa \delta \,  (\Lambda/\kappa)^{2/3}$, 
the integral becomes independent of $\delta$ as is seen by using $k = p e^{\kappa \tau}/ (\Lambda^2\kappa)^{1/3}$. 
Explicitly one finds
\be
\partial_x \partial_{x_0} G^{in}(x,0 ; x_0, -\tau) \sim  (\Lambda^2 \kappa)^{2/3} e^{-\kappa \tau} 
\times C_\pm(\kappa/\Lambda,\tau),
\label{feq}
\ee
where $C_\pm(\kappa/\Lambda,\tau)$ are slowly varying functions which stay bounded for $\tau \to \infty$.
This exponentially decreasing result can be seen as the contribution on the horizon
of the tail of the configurations with high $p$ which follow the second equation in \eqr{modchar}. 

This smoothing out of the relativistic behavior is very reminiscent to what was found in~\cite{Barrabes:2000fr}
when studying the backwards evolution of $ G^{in}(x,0 ; x_0, -\tau)$ of a {\it relativistic} field 
propagating in a stochastically fluctuating black hole metric. 
In addition, for nearby points, $ G^{in}$ in a stochastic geometry also behaved as \eqr{Geqt}, 
as can be seen in Eq. (4.7).
Based on this similarity it was argued~\cite{Parentani:2007mb} that when taking into account 
the gravitational radiative corrections,  
the dressed Green functions should
effectively behave near a black hole horizon as 
in \eqr{feq}, thereby 
reinforcing the
idea 
that the unbounded growth of \eqr{ddcfd} cannot "accommodate gravitational non-linearities".

\section{Conclusions}

We showed that the monotonic energy correlations found in the vacuum \eqr{TUU}
gives rise to a maximum of correlation across a Rindler horizon 
when re-expressed in terms of coordinates associated with accelerated systems, see \eqr{ldTuu}. 
This maximum is not a mere coordinate artefact as it affects the combined state
of co-accelerating systems. 

In Sec. III, we transposed this analysis to stationary black hole geometries, and recalled
that the regularity of the state across the horizon and the inertial character of
asymptotic observers are essential to provide a physical meaning to the thermal
correlations of \eqr{Tuu}. When considering  black hole geometries which contain
asymptotic regions on both sides of the horizons, the correlations of 
\eqr{ldTuu} are found at large distances when using inertial coordinates. 
We then make use of the affine parameter $r_v$ of \eqr{rv} to obtain 
an invariant description of the energy correlations in the entire space-time.
We compared the correlation pattern associated with a late detection, Fig. 1, to that
obtained at equal EF time, Fig. 2. 
In both cases the gradual emergence of a maximum of correlations across the horizon
is clearly visible.
 By considering the subtracted 
correlations of \eqr{Sr}, we saw that the remaining signal is dominated by the
long distance correlations across the horizon, and also contains a
sub-dominant local contribution associated with \eqr{Tuu}.
This analysis was generalized in III.F. by including the transients effects which
precede the stationary patterns found in the Unruh vacuum.

In Sec. IV. by studying both wave packets and correlation functions,
we studied 
 how these patterns are modified by dispersion. 
Far away from the horizon, the pattern is robust, i.e. hardly affected by dispersion. 
Close to the horizon we saw that dispersive effects show up differently depending on how
one probes the state. When probed at equal PG time, the correlation function is translation
invariant, and as in Minkowski, see \eqr{Geqt}. Instead wave packets of $in$ modes
centered around a given frequency $\om$ 
display a characteristic pattern which follows the modified characteristics of \eqr{modchar}. 
When the momentum has sufficiently increased (in a backward in time propagation) 
the wave packets are dragged away from the horizon, 
and, the blue shift effect saturates. 
This behavior is recovered from the correlation function when considered at different times
and appears through a non-trivial saddle point in 
\eqr{mG}.
When considering the correlation function for two points separated by a very large PG time,
the drag w.r.t. the relativistic horizon results in an exponentially suppressed amplitude 
in the place of the exponentially growing result found in relativistic theories, compare
\eqr{ddcfd} with \eqr{mGd}. These properties are reminiscent to what was 
found 
when considering field propagation in a 
fluctuating black hole metric, and could possibly be found when taking into
account gravitational interactions at the quantum level.

\begin{acknowledgments}
We would like to thank T. Jacobson for discussions concerning the affinity of $r$,
and R. Balbinot, S. Finazzi and N. Obadia for useful remarks.
\end{acknowledgments}

\appendix

\section{Unruh modes}

Firstly, they are solutions of d'Alembert equation $\partial_U\partial_V \phi_\om = 0$, and
thus only depend on either $U$ or $V$. Secondly, they have a fixed boost frequency $\om$,
i.e. they are eigenmodes of 
\be
i\partial_u \phi_\om = - i aU\partial_U \phi_\om = \om \phi_\om .
\label{eigm}
\ee
Thirdly they are only composed of the 
positive norm 
 modes: $\phi_\Om = e^{-i \Om U}/(4 \pi \Om)^{1/2}$ with $\Om > 0$. Explicity they are given by
\ba
\phi_\om &=& \frac{\alpha_\om}{(4\pi \om)^{1/2}} \, \left(- aU + i \epsilon \right)^{i\om/a}
\nonumber \\
&=& 
\frac{\alpha_\om}{(4\pi \om)^{1/2}}
\Big[ \theta(-U) (-aU)^{i\om/a} + z_\om \times \theta(U) (aU)^{i\om/a} \Big] ,
\label{Um}
\ea
where the normalization 
obeys 
$ \vert \alpha_\om\vert^2 = 
(1 - e^{- 2\pi \om/a})^{-1}$, 
and where 
 $z_\om = e^{-\pi\om /a}$. This factor arises 
from the $i\epsilon$ prescription which specifies that the 
analytic continuation from the $R$ to $L$ quadrant
 must be done in {\it lower} half 
complex $U$ plane. 
%
%
As in \eqr{TUU} this prescription comes from the fact that
only positive frequency $\Om = i\partial_U$ modes 
contribute.


Moreover, they are globally defined, $- \infty < U < \infty $, and form a complete and orthonormal   
basis of positive norm modes when $- \infty < \om< \infty $ (with respect to the standard Klein-Gordon product).  
Hence the Minkowski vacuum can be 
alternatively 
defined as the state 
annihilated by the destruction operators $a_\om$ associated with these modes. 

Thus, when a quantum system is (linearly) coupled to $\phi$ which is initially 
in the vacuum, 
the 
transition amplitudes 
will contain some (linear) combination of the 
$\phi_\om$. 
When the 
system is not accelerated (e.g. inertial),
the decomposition (\ref{Um}) presents no interest since 
the system will cross $U=0$.
On the contrary, when it 
is uniformly accelerated
in, say, the $R$ quadrant, \eqr{Um}
guarantees that {\it every} transition occurring in the Minkowski vacuum
defines 
a partner wave 
in $L$, see App. C. for more details. 

This $R-L$ partnership 
can be studied 
in simpler terms and without referring 
to accelerated systems
by 
constructing wave packets of Unruh modes
\be
\bar \phi = \int_0^\infty d\om \bar f_\om \phi_\om = \bar \phi^R + \bar \phi^L.
\label{wp}
\ee
\eqr{Um} thus implies that to {every} packet $\bar \phi^R$ localized in $R$ will correspond $\bar \phi^L$, 
its partner wave  in $L$.
More can be said: since 
$z_\om$ in the r.h.s. is real for all $\om$, 
when $\bar \phi^R$ 
constructively interferes 
around some $\bar U_R < 0$, 
\eqr{Um} guarantees that $\bar \phi^L$
will do so 
near $-\bar U_R$. This explains why the maximum in \eqr{ldTuu} arises for opposite 
values of 
$U$. In addition, from the fact that 
high $\om$ 
are exponentially suppressed by $z_\om= e^{-\pi \om/a}$, 
the maximum 
in \eqr{ldTuu} cannot diverge as it does in \eqr{Tuu}. Instead it must scale as $a^4$. 
In brief, the mathematical properties of \eqr{ldTuu} are deeply rooted to those
of the modes  $\phi_\om$. 


There exists an efficient way to encode the properties of the Unruh modes 
which turns out to be very useful when analyzing Hawking radiation 
in the presence of dispersion. It consists in computing the 
Fourier transform  at fixed $t$: 
 $\tilde \phi_\om(p) = \int dz \, e^{- ipz} \phi_\om/(2 \pi)^{1/2}$.
Taking into account the $i \epsilon$ in \eqr{Um}, up to an irrelevant phase, one finds
\be
\tilde \phi_\om(p) = \theta(p) \, {p^{- i\om/a - 1}\over ( 4 \pi a)^{1/2}}.
\label{Ump}
 \ee
The restriction to positive $p$ follows from the fact that property only positive $\Om$ contribute to $\phi_\om$,
and from the dispersion relation
$\Om= p$ which 
describes right moving modes.
We also note that when considering 
\eqr{wp} in the $p$-representation,
the two wave packets found in the $x$-representation on either side the horizon 
are now described by a {\it single} packet in $p$-space. This is characteristic of 
pair production phenomena, see e.g. Sec. 1.2-1.3 in \cite{Brout:1995rd}.
We shall return to this in Sec. IV.


\section{Fulling-Rindler states and partner-ship in Fock space}


If a mode analysis is sufficient to understand the behavior of \eqr{ldTuu} in space-time,
to have a deeper quantum mechanical understanding of \eqr{ldTuu} we analyze 
the bi-partite structure in the Fock space when using states with a fixed frequency $\om$.
To this end, we should discuss yet another property of 
\eqr{Um}. It concerns the fact that 
the mode on the left ($U> 0$) has a negative norm (for $\om > 0$) 
thereby implying that the norm of the right component is
correspondingly larger. 
This 
invites to consider the inequivalent quantization of $\phi$
based on the "Fulling-Rindler" (FR) modes. 
These are normalized eigenmodes of frequency $\om$, \eqr{eigm}, 
localyzed either on the right, or the left, of $U=0$. 
Hence we re-write \eqr{Um} as 
\ba
\phi_\om 
&=& \alpha_\om\,  
\phi^R_\om
+ \beta_\om 
 \left(
\phi^L_\om 
\right)^*  , \quad {\rm for} \ \om > 0,
\nonumber\\ 
\phi_\om &=& \alpha_{-\om}\,  
\phi^L_{-\om}
+ \beta_{-\om} 
 \left(
\phi^R_{-\om} 
\right)^* ,\quad {\rm for} \ \om <  0,
\label{Bo}
\ea
where $\phi^R_\om =  
{e^{-i\om u}}/{(4\pi \om)^{1/2}}$ 
($\phi^L_\om = 
{e^{-i\om \bar u}}/{(4\pi \om)^{1/2}}$) 
vanishes on the left (right) 
of the horizon,
and where $\beta_\om = z_\om \alpha_\om$. 
One easily verifies that $\alpha_\om^2 - \beta_\om^2 = 1$, which implies
$\beta_\om^2 
= (e^{2\pi \om/a} - 1)^{-1}$.
%
%
For each $\om > 0$, 
\eqr{Bo} defines a (two-mode) 
Bogoliubov transformation relating 
$(\phi_\om,\phi_{-\om})$
to 
$(\phi^R_\om,\phi^L_\om)$. This implies that 
the vacuum can be written as a product over $\om > 0$ of two-mode squeezed states 
\be
|0\rangle = \Pi_\om \left(\frac{1}{\alpha_\om}\exp{(z_\om \,  a^R_\om a^L_\om)^\dagger}\right)\,  |0\rangle_R \,  |0\rangle_L , 
\label{sqs}
\ee
where the $R$-vacuum $|0\rangle_R $ is the state annihilated by the $a^R_\om$, 
the destruction operators associated with the 
$\phi_\om^R$, and similarly for the $L$ sector.
Since the squeezing operator is quadratic and diagonal in $\om$, for free fields,
{\it all} expectation values are expressible in terms of the following two VEV
\ba
\langle (a^R_\om)^\dagger  a^R_\om \rangle &=& \langle (a^L_\om)^\dagger  a^L_\om \rangle = \vert \beta_\om\vert^2 ,  
\label{diag} \\
\langle a^R_\om  a^L_\om \rangle &=&   \beta_\om \alpha_\om^* = z_\om \, \vert \alpha_\om\vert^2 .  
\label{ndiag}
\ea

It is now instructive to see how these two VEV enter in \eqr{Tuu} and \eqr{ldTuu}.
To this end, we consider 
the ($U$ contribution of the) 2 point function of $\phi$ 
\be
\langle \phi({U}) \,  \phi({U_0})\rangle = \int^\infty_{-\infty}d\om\,  \phi_\om({U}) (\phi_\om({U_0}))^*
= - \frac{1}{4\pi} \ln({U} - {U_0} - i \epsilon).
\label{2ptphi}
\ee
Using \eqr{Bo}
 two different expressions are obtained depending 
if 
both points are on one side, or on either side, 
of $U= 0$. 
Explicitely, when both $U$ are negative and written as $-aU = e^{-au}$ 
one has%
\ba
\langle \phi({U}) \,  \phi({U_0})\rangle &=& \int^\infty_{0}\frac{ d\om}{4\pi \om}\, 
\left(| \alpha_\om|^2 e^{-i\om({u} - {u_0})} + |\beta_\om|^2 e^{+ i\om({u} - {u_0})}\right).
\label{bUneg}
\ea
Instead when one point, say $U$, is positive and written as $aU=e^{a\bar u}$, one has
\ba
\langle \phi(U) \,  \phi({U_0})\rangle &=& 
\int^\infty_{0}\frac{ d\om}{4\pi \om}\, 2 {\rm Re}
\left( \vert \alpha_\om\vert^2  z_\om^* \, \, e^{-i\om(\bar {u} + {u_0})}
\right).
\label{Unp}
\ea
%
One sees 
that \eqr{bUneg} and \eqr{Tuu} arise from diagonal terms, 
hence weighted by $|\beta_\om|^2$ of \eqr{diag}, 
whereas 
\eqr{Unp} and \eqr{ldTuu} arise from interfering terms weighted by $\alpha_\om^* \beta_\om$
of \eqr{ndiag} which encodes the entanglement, {\it in Fock space}, amongst the $R$ and $L$ sectors.

It is an interesting exercice to verify that when using the actual functions for $\alpha_\om$ and $\beta_\om$, 
\eqr{bUneg} and \eqr{Unp} both give back, as they must, the $\log \Delta U$ 
of \eqr{2ptphi}. Therefore, 
they are only complicated re-expressions of \eqr{2ptphi}.
Similarly, 
\eqr{sqs} is only a mathematical re-expression of the Minkowski vacuum.
However, it prepares the analysis of the physical processes related to the Unruh effect, 
to the quantum fluxes emitted by non-uniform mirrors~\cite{Carlitz:1986ng,Obadia:2002qe}, and to black hole physics.
In these three cases, there is an external agent --respectively an accelerated system, a non-uniform mirror,
a non-trivial metric-- which acts on the field and "transforms" the FR states into asymptotic states.
This particularly neat in the case of the non-uniform mirror discussed in \cite{Carlitz:1986ng},
see Eqs. (3.27-3.28), see also Section 2.5 in \cite{Brout:1995rd}. 


\section{The conditional value associated with a detection in $R$}

We recall how a detection of a FR quantum in $R$
 defines first, a partner state in $L$, and second, 
a projector which allows to define the conditional value of an operator associated with this detection. 
We describe the detected quantum in $R$ by
\be
\vert \bar \Psi_R \rangle = \int_0^\infty d\om f_\om \, (a^R_\om)^\dagger \vert 0 \rangle_R .
\ee
The EPR partner state is defined by reducing the bi-partite state.
In the present case, the latter is the Minkowski vacuum expressed as \eqr{sqs}. 
The partner state is thus
\be
\vert \bar \Psi_L \rangle = \langle \bar \Psi_R \vert 0 \rangle = 
\int_0^\infty d\om z_\om^* \, f^*_\om \, (a^L_\om)^\dagger \vert 0 \rangle_L .
\ee
One notices that its Fourier components are fixed by  $ z_\om^*$ and $ f^*_\om$, i.e. by both 
the state and the complex conjugated of the component of selected wave packet. 

It is instructive to relate these two states to the (classical) wave packets 
of 
\eqr{wp}. 
To this end we introduce the projector $\Pi_R = \vert \bar \Psi_R  \rangle\langle \bar \Psi_R \vert$
and consider the value of $\phi^2$ conditional to the fact that the detection took place 
(for more details see \cite{Brout:1995rd}) 
\be
\bar \phi^2 = \langle 0 \vert \, \phi^2 \, \Pi_R \, \vert 0 \rangle = (\bar \phi^2)^R + (\bar \phi^2)^L .
\ee
A direct calculation gives
\ba
(\bar \phi^2)^R &=& \left( \int_0^\infty d\om f_\om \phi_\om^R \right)
 \left( \int_0^\infty d\om f_\om \vert z_\om \vert^2\phi_\om^R \right)^* , \nonumber \\
(\bar \phi^2)^L &=& \vert \int_0^\infty d\om f_\om^* z_\om^* \phi_\om^L \vert^2 .
\ea
When $f_\om=\alpha_\om \bar f_\om$ where $\bar f_\om$ given in \eqr{wp}, 
$(\bar \phi^2)^L$
 exactly gives $\vert \bar \phi_L \vert^2$ of that equation.
Similarly the first factor in the first line is $\bar \phi_R$. The second factor is 
not its complex conjugated due to the presence of $\vert z_\om \vert^2$ in the integrand.
However for wave packets with a small spread $\om$ wrt $a$, this quantum mechanical feature (whose consequences
are discussed in \cite{Massar:1996tx})
does not significantly affect the spatial properties of the $R$ wave packet. 
Thus we basically recover the modulus square  of the two packets of \eqr{wp}. 

The lesson of this exercise is that the pattern obtained by constructing 
wave packets of Unruh modes as in \eqr{wp} offers a reliable description 
of the quantum correlations across a Rindler horizon
(when the spread in $\om$ is small enough).
This transposes in black hole metrics (without and with dispersion) and implies that the 
correlation patterns of $in$ modes also offer a reliable description 
of the quantum correlations across the horizon.

Finally, we mention that the study of highly excited coherent states, see App. C of \cite{Macher:2009nz},
offers another way to relate the packets of \eqr{wp}
 to quantum states. 
Using these coherent states, one can verify 
the agreement of both descriptions in describing the $R-L$ correlations.  





\begin{thebibliography}{99}



\bibitem{Unruh:1976db}
  W.~G.~Unruh,
  Phys.\ Rev.\  D {\bf 14}, 870 (1976).

\bibitem{Unruh:1983ms}
  W.~G.~Unruh and R.~M.~Wald,
  Phys.\ Rev.\  D {\bf 29}, 1047 (1984).


\bibitem{Carlitz:1986ng}
  R.~D.~Carlitz and R.~S.~Willey,
  Phys.\ Rev.\  D {\bf 36}, 2327 (1987),
and 2336 (1987).

\bibitem{Massar:1995im}
  S.~Massar and R.~Parentani,
  Phys.\ Rev.\  D {\bf 54}, 7426 (1996).

\bibitem{Brout:1995rd}
  R.~Brout, S.~Massar, R.~Parentani and Ph.~Spindel,
  Phys.\ Rept.\  {\bf 260}, 329 (1995).

\bibitem{Obadia:2002qe}
  N.~Obadia and R.~Parentani,
  Phys.\ Rev.\  D {\bf 67}, 024022 (2003), and
024021 (2003).

\bibitem{Massar:2006ev}
  S.~Massar and P.~Spindel,
  Phys.\ Rev.\  D {\bf 74}, 085031 (2006).


\bibitem{Hawking:1974sw}
  S.~W.~Hawking,
  Commun.\ Math.\ Phys.\  {\bf 43}, 199 (1975)
  [Erratum-ibid.\  {\bf 46}, 206 (1976)].


\bibitem{Davies:1976ei}
  P.~C.~W.~Davies, S.~A.~Fulling and W.~G.~Unruh,
  Phys.\ Rev.\  D {\bf 13}, 2720 (1976).

\bibitem{Massar:1996tx}
  S.~Massar and R.~Parentani,
  Phys.\ Rev.\  D {\bf 54}, 7444 (1996), 
  {}F.~Englert, S.~Massar and R.~Parentani,
{}Class.\ Quant.\ Grav.\  {\bf 11}, 2919 (1994). 

\bibitem{Unruh:1994je}
  W.~G.~Unruh,
  Phys.\ Rev.\  D {\bf 51}, 2827 (1995).

\bibitem{Unruh:1980cg}
  W.~G.~Unruh,
  Phys.\ Rev.\ Lett.\  {\bf 46}, 1351 (1981).

\bibitem{Barcelo} C.~Barcelo, S.~Liberati, and M.~Visser,
Living Rev. Relativity {\bf 8}, 12 (2005).



\bibitem{Jacobson:1996zs}
  T.~Jacobson,
  Phys.\ Rev.\  D {\bf 53}, 7082 (1996).

\bibitem{Schutzhold:2010ig}
  R.~Schutzhold and W.~G.~Unruh,
  arXiv:1002.1844 [gr-qc].


\bibitem{Brout:1995wp}
  R.~Brout, S.~Massar, R.~Parentani and P.~Spindel,
  Phys.\ Rev.\  D {\bf 52}, 4559 (1995).


\bibitem{Corley96}
S. Corley and T. Jacobson, Phys. Rev. D {\bf 54}, 1568 (1996).



\bibitem{Balbinot:2007de}
  R.~Balbinot, A.~Fabbri, S.~Fagnocchi, A.~Recati and I.~Carusotto,
  Phys.\ Rev.\  A {\bf 78}, 021603 (2008). 


\bibitem{Iacopo08}
  I.~Carusotto, S.~Fagnocchi, A.~Recati, R.~Balbinot and A.~Fabbri,
  New J.\ Phys.\  {\bf 10}, 103001 (2008).

\bibitem{Macher:2009nz}
  J.~Macher and R.~Parentani,
Phys.\ Rev.\  A {\bf 80}, 043601 (2009).   



\bibitem{Barrabes:2000fr}
C.~Barrabes, V.~P.~Frolov and R.~Parentani,
{}Phys.\ Rev.\  D {\bf 62}, 044020 (2000).

\bibitem{Parentani:2007mb} 
{}R.~Parentani,
  {}Int.\ J.\ Theor.\ Phys.\  {\bf 41}, 2175 (2002)
  [arXiv:0704.2563], and
{}Phys.\ Rev.\  D {\bf 63}, 041503 (2001).



\bibitem{Jacobson:1995ab}
  T.~Jacobson,
  Phys.\ Rev.\ Lett.\  {\bf 75}, 1260 (1995).

\bibitem{Damour:1976jd}
  T.~Damour and R.~Ruffini,
  Phys.\ Rev.\  D {\bf 14}, 332 (1976).

\bibitem{Unruh1977}
  W.~G.~Unruh,
  Phys.\ Rev.\  D {\bf 15}, 365 (1977).

\bibitem{Campo:2003pa}
  {}D.~Campo and R.~Parentani, {}Phys.\ Rev.\  D {\bf 70}, 105020 (2004).

\bibitem{Parentani:1994ij}
  R.~Parentani and T.~Piran, 
{}Phys.\ Rev.\ Lett.\  {\bf 73}, 2805 (1994).

\bibitem{Jacobson:1991gr}
  T.~Jacobson,
  Phys.\ Rev.\  D {\bf 44}, 1731 (1991), D {\bf 48}, 728 (1993).

\bibitem{Parentani:1993yz} 
  R.~Parentani, Class.\ Quant.\ Grav.\  {\bf 10}, 1409 (1993). 


\bibitem{Jacobson:2007jx}
  T.~Jacobson and R.~Parentani,
  Phys.\ Rev.\  D {\bf 76}, 024006 (2007).

\bibitem{Jacobson:1999zk}
  T.~Jacobson,
  Prog.\ Theor.\ Phys.\ Suppl.\  {\bf 136}, 1 (1999)
  [arXiv:hep-th/0001085].


\bibitem{Jacobson:2008cx} T.~Jacobson and R.~Parentani, {}Class.\ Quant.\ Grav.\  {\bf 25}, 195009 (2008).


\bibitem{Balbinot:2006ua}
  R.~Balbinot, A.~Fabbri, S.~Fagnocchi and R.~Parentani,
  Riv.\ Nuovo Cim.\  {\bf 28}, 1 (2005)
  [arXiv:gr-qc/0601079].


\bibitem{Corley}   
 S. Corley, Phys. Rev. D {\bf 57}, 6280 (1998). 


\bibitem{Macher:2009tw}
  J.~Macher and R.~Parentani,
  Phys.\ Rev.\  D {\bf 79}, 124008 (2009). 




\bibitem{Balbinot:2007kr}
  {}R.~Balbinot, A.~Fabbri, S.~Farese and R.~Parentani,
  {}Phys.\ Rev.\  D {\bf 76}, 124010 (2007).



\bibitem{Bombelli:1986rw}
        L.~Bombelli, R.~K.~Koul, J.~H.~Lee and R.~D.~Sorkin,
        Phys.\ Rev.\  D {\bf 34}, 373 (1986).




\bibitem{Campo:2005sv}
{}D.~Campo and R.~Parentani, 
  {}Phys.\ Rev.\  D {\bf 74}, 025001 (2006).

\end{thebibliography}
\end{document}